\documentclass[twocolumn,english,aps,prb,floats]{revtex4-2}
\usepackage[T1]{fontenc}
\usepackage[latin9]{inputenc}
\usepackage{babel}
\usepackage{amsmath}
\usepackage{amssymb}
\usepackage{graphicx}
\usepackage{esint}
\usepackage[unicode=true,
 bookmarks=false,
 breaklinks=false,pdfborder={0 0 1},backref=section,colorlinks=false]
 {hyperref}

\makeatletter

\newcommand{\lyxdot}{.}

\usepackage{babel}

\@ifundefined{textcolor}{}{%
 \definecolor{BLACK}{gray}{0}
 \definecolor{WHITE}{gray}{1}
 \definecolor{RED}{rgb}{1,0,0}
 \definecolor{GREEN}{rgb}{0,1,0}
 \definecolor{BLUE}{rgb}{0,0,1}
 \definecolor{CYAN}{cmyk}{1,0,0,0}
 \definecolor{MAGENTA}{cmyk}{0,1,0,0}
 \definecolor{YELLOW}{cmyk}{0,0,1,0}
}

\usepackage{ifpdf}\usepackage{bm}

\makeatother

\begin{document}
\title{Skyrmion Crystal in a Microwave Field}
\author{Dmitry A. Garanin and Eugene M. Chudnovsky}
\affiliation{Physics Department, Herbert H. Lehman College and Graduate School,
The City University of New York, 250 Bedford Park Boulevard West,
Bronx, New York 10468-1589, USA }
\date{\today}
\begin{abstract}
Temperature and field dependences of the frequencies of uniform modes
of the skyrmion lattice in a 2D ferromagnetic film with Dzyaloshinskii-Moriya
interaction, as well as their damping, are computed within the model
of classical spins. We show that the magnetization of the film exhibits
Rabi-like oscillations when subjected to the microwave field at resonance
with the low-frequency mode. Melting of the skyrmion lattice by resonant
microwaves is investigated in terms of the time dependence of the
orientational and translational order parameters. A distinct single-stage
melting transition has been observed.
\end{abstract}
\maketitle

\section{Introduction}

\label{Sec_Introduction}

Studies of topological defects in solids, such as skyrmions \cite{SkyrmePRC58,Polyakov-book,Manton-book},
have been inspired by the beauty of physics and mathematics associated
with them, and more recently by their potential for developing topologically
protected information technology \cite{Nagaosa2013,Zhang2015,Klaui2016,Hoffmann-PhysRep2017,Fert-Nature2017,Chen-Nature2024}.
In magnetic films, they can be stabilized by Dzyaloshinskii-Moriya
interaction (DMI) \cite{Bogdanov1989,Bogdanov94,Bogdanov-Nature2006,Heinze-Nature2011,Boulle-NatNano2016,Leonov-NJP2016},
frustrated exchange interactions \cite{Leonov-NatCom2015,Zhang-NatCom2017},
magnetic anisotropy \cite{IvanovPRB06,Lin-PRB2016}, disorder \cite{CG-NJP2018},
or geometrical confinement \cite{Moutafis-PRB2009}. Within a certain
range of parameters, skyrmions form regular periodic lattices. They
have been first predicted theoretically \cite{Bogdanov94} and subsequently
observed in experiments, initially in FeCoSi films \cite{Yu2010}
and then in many other materials, which prompted research on phase
diagrams of the skyrmion matter, see, e.g., Refs.\ \cite{GCZZ-MMM2021,Dohi-ARCMP2022,DG-EC-JMMM2024}
and references therein.

\begin{figure}
\centering{}\includegraphics[width=8cm]{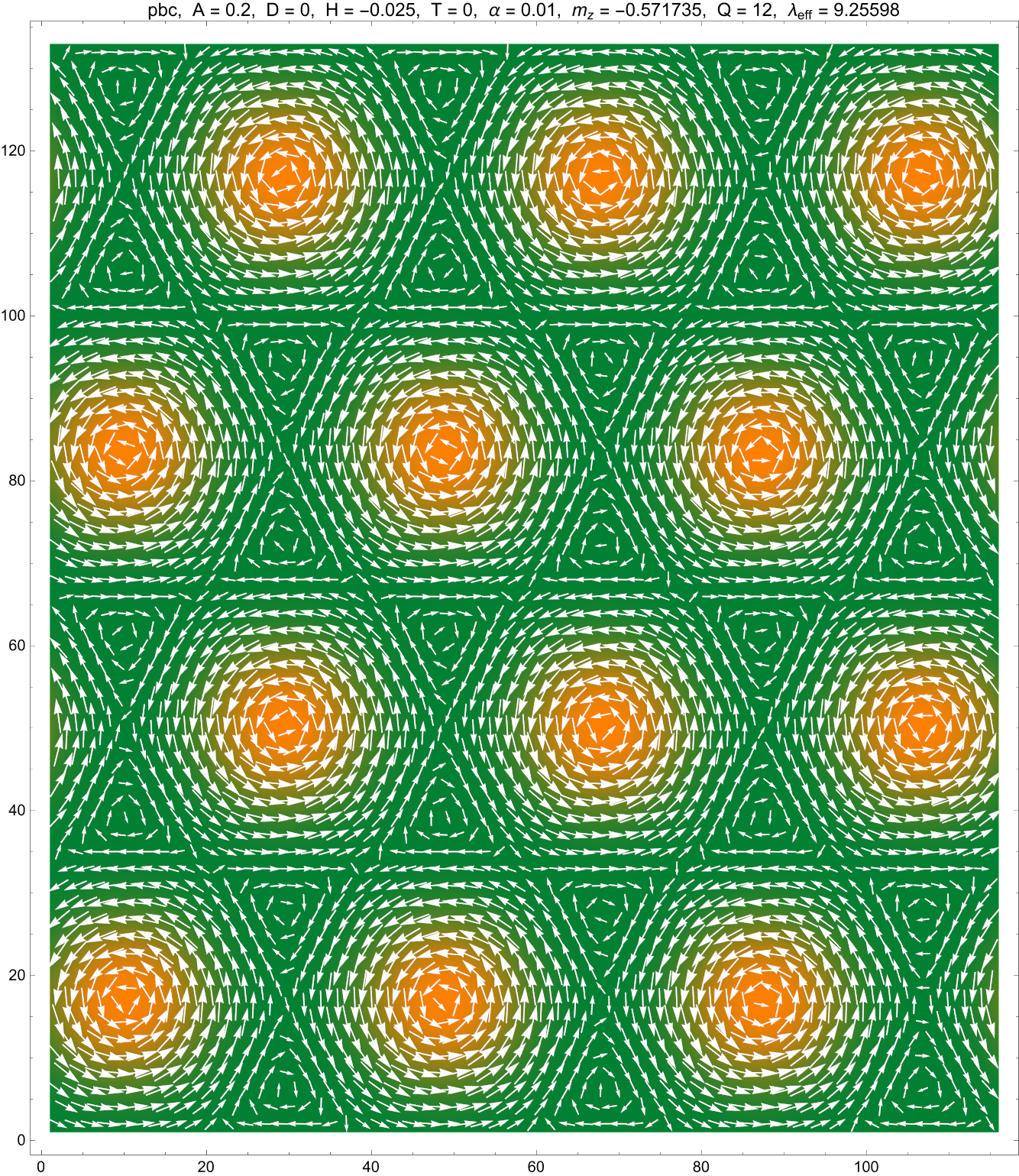} \caption{Hexagonal skyrmion lattice at $T=0$ obtained by the relaxation of
the initial state of topological bubbles defining the topological
charge $Q$ of the system. Spin components are color coded: $s_{z}=-1$
green, $s_{z}=1$ orange. White arrows show the in-plane spin components
$s_{x}$ and $s_{y}$. The latter are directed in the circular direction
(depending on the sign of $A$) around skyrmion centers for the Bloch-type
DMI which was used in Eq. (\ref{Ham}). For the Néel-type DMI, $s_{x}$
and $s_{y}$ are directed in the radial direction away or toward the
skyrmions centers, depending on the sigh of $A$.}
\label{Fig_SkL} 
\end{figure}

More recently, people began investigating microwave (MW) properties
of skyrmion lattices (SkL), see Fig.\ \ref{Fig_SkL}. Such experiments
are relatively easy to perform and compare with theoretical predictions.
Excitation modes of individual skyrmions typically lie in the same
(microwave) frequency range as the frequency of the ferromagnetic
resonance (FMR), see, e.g., Ref.\ \cite{DG-RJ-EC-PRB2020} where
skyrmion breathing mode was investigated, and references therein.
Mochizuki \cite{Mochizuki-PRL2012} studied excitations of the SkL
within a two-dimensional (2D) spin model using the Landau-Lifshitz
equation. He found two other, precessing, modes in addition to the
skyrmion breathing mode, and observed melting of the skyrmion crystal
by microwaves. Onose et al. \cite{Onose-PRL2012} subsequently observed
these modes in a helimagnetic insulator Cu$_{2}$OSeO$_{3}$. These
findings were later confirmed by Aqeel et al. \cite{Aqeel-PRL2021}
and Lee et al. \cite{Lee-JPhys2022} who also reported evidence of
the hybridization of precessing and breathing modes. MW resonances
of magnetic skyrmions in thin-film multilayers were investigated experimentally
by Satywali et al. \cite{Satywali-NatCom2021}. Li et al. \cite{Li-JPhys2023}
studied the collapse of the SkL induced by microwaves within a micromagnetic
model.

In this article, we address questions that have not received much
attention in the past. The temperature and magnetic field dependence
of the uniform modes of a 2D hexagonal SkL stabilized by the Heisenberg
exchange, DMI, and the magnetic field has been studied, and the damping
of the modes due to spin-wave processes caused by nonlinearity has
been computed. Skyrmion lattices provided a rich novel system for
the study of 2D melting \cite{Ambrose,Nishikawa-PRB2019,Zazvorka2020,Huang-Nat2020,Balaz-PRB2021,McCray2022,GC-PRB2023,Meisenheimer2023,DG-EC-JMMM2024,DG-JS-EC-JPhys2024}
-- the problem that was packed with controversies over the last 50
years, see, e.g., Refs.\ \cite{Kosterlitz-Review2016,Anderson2017,Ryzhov2023}
for review. It was eventually established that the originally proposed
two-stage melting scenario \cite{HN-PRL1978,NH-PRB1979,Young-PRB1979}
was not universal but depended on the energy of the dislocation core,
the interaction potential, the shape of the interacting particles,
and the symmetry of the crystal lattice. Melting of the SkL by MW
provides another dimension to this problem. We compute the behavior
of the orientational and translational order parameters and show that
this kind of melting is a one-stage process similar to the conventional
melting of a solid.

The paper is organized as follows. The model of the SkL and the order
parameters are introduced in Sec.\ \ref{Sec_model_and_theory}. The
numerical method is described in Sec.\ \ref{Sec_numerical_method}.
MW resonances, their damping, and their temperature and field dependence
are computed and in Sec.\ \ref{Sec_FDT}. The melting of the SkL
on continuously pumping the MW power into the system is studied in
Sec.\ \ref{Sec_MW}. Our conclusions and suggestions for experiments
are presented in Sec.\ \ref{Sec_conclusions}.

\section{The model and theoretical approaches}

\label{Sec_model_and_theory}

\subsection{The model}

\label{Sec_model}

We study skyrmion phases of ferromagnetically coupled classical three-component
spins ${\bf s}_{i}$ of length $1$ on a square lattice. The Hamiltonian
of the system is 
\begin{eqnarray}
{\cal H} & = & -\frac{1}{2}\sum_{ij}J_{ij}{\bf s}_{i}\cdot{\bf s}_{j}-A\sum_{i}\left[({\bf s}_{i}\times{\bf s}_{i+\delta_{x}})_{x}+({\bf s}_{i}\times{\bf s}_{i+\delta_{y}})_{y}\right]\nonumber \\
 & - & H\sum_{i}s_{iz}-{\bf h}(t)\cdot\sum_{i}{\bf s}_{i}.\label{Ham}
\end{eqnarray}
The first term is the ferromagnetic exchange interaction of strength
$J>0$ between neighboring spins. The second term is the Bloch-type
DMI interaction of strength $A$, with $\delta_{x}$ and $\delta_{y}$
denoting the nearest lattice sites in the positive $x$ or $y$ direction.
(The Néel-type DMI has $({\bf s}_{i}\times{\bf s}_{i+\delta_{x}})_{y}+({\bf s}_{i}\times{\bf s}_{i+\delta_{y}})_{x}$
instead. It produces similar results.) The third term in Eq.\ (\ref{Ham})
is the Zeeman interaction with the external field $H<0$ applied in
the negative $z$-direction, and the last term is the Zeeman interaction
with the ac field ${\bf h}(t)$ of arbitrary polarization.

A three-component two-dimensional spin field possesses a topological
charge 
\begin{equation}
Q=\frac{1}{4\pi}\int dxdy\,{\bf s}\cdot\left(\frac{\partial{\bf s}}{\partial x}\times\frac{\partial{\bf s}}{\partial y}\right)\label{Q_def}
\end{equation}
that takes discrete values $Q=0,\pm1,\pm2,...$ For the pure-exchange
model, the analytical solution for topological configurations with
a given value of $Q$ was found by Belavin and Polyakov \cite{belpol75jetpl}
(see also Ref. \cite{capgarchu20jpcm}). The value of $Q$ shows how
many times the spin vector, rotating in space, circumscribes the full
body angle $4\pi$. Due to the scale invariance of the exchange interaction
in 2D, their energy with respect to the uniform state does not depend
on their size $\lambda$ and for the structures with $Q=\pm1$, skyrmions
and antiskyrmions, equals to $\Delta E_{\mathrm{BP}}=4\pi J$ with
respect to the energy of the uniform state. Skyrmions oriented up
($s_{z}=1$ at the center of the skyrmion and $s_{z}=-1$ far away
from the center, i.e., polarity $p=1$ ) have $Q=1$. Rotating all
spins by 180$^{\circ}$around an axis within the $xy$ plane results
in a down-oriented antiskyrmion ($p=-1$) and the same $Q$. Reversing
all spins results in a skyrmion with $p=-1$ and $Q=-1$. For skyrmions,
polarity and topological charge coincide while for antiskyrmions they
are opposite \cite{kovsan18front}. Using the formalism by Belavin
and Polyakov, one can construct analytical solutions for different
kinds of skyrmion lattices \cite{capgarchu19prb}.

Finite lattice spacing $a$ breaks this invariance by adding a term
of the order $-\left(a/\lambda\right)^{2}$ to the energy, which leads
to the skyrmion collapse \cite{caichugar12prb}. Equation (\ref{Q_def})
can be generalized for discrete spins on the lattice using the formula
for the body angle circumscribed by a triad of vectors \cite{eriksson90mm}.
The result has the form
\begin{equation}
Q=\frac{1}{2\pi}\sum_{i,\epsilon=\pm1}\arctan\frac{\mathbf{s}_{i}\cdot\left(\mathbf{s}_{j}\times\mathbf{s}_{k}\right)}{1+\mathbf{s}_{i}\cdot\mathbf{s}_{j}+\mathbf{s}_{j}\cdot\mathbf{s}_{k}+\mathbf{s}_{k}\cdot\mathbf{s}_{i}},\label{Q_def_discrete}
\end{equation}
where $j\equiv i+\epsilon\delta_{x}$ and $k\equiv i+\epsilon\delta_{y}$.
In the continuous limit, the numerator of this formula becomes small
and the cross-product can be approximated by derivatives, whereas
the denominator approaches the value of four, so that arctan can be
discarded. Taking into account the factor 2 from the summation over
$\epsilon$ and replacing summation by integration, one recovers Eq.
(\ref{Q_def}). At elevated temperatures, neighboring spins deviate
by substantial angles from each other so that the continuous approximation
breaks down and Eq. (\ref{Q_def}) yields smaller values of $Q$.
To the contrast, Eq. (\ref{Q_def_discrete}) is robust and remains
valid at elevated temperatures.

In the presence of other interactions, solutions for skyrmions and
skyrmion lattices can be obtained numerically. Stable solutions exist
in a certain range of $A/J$ and $H/J$. In this paper we use $A/J=0.2$
which is typical for materials with nanoscale skyrmions \cite{Leonov-NJP2016,Camley2023}
with $H/J=-0.025$ as the main case. For these parameters, the energy
of SkL is minimal at the distance between the neighboring skyrmions
$a_{S}=38.5a$ \cite{GC-PRB2023} which is the main case we consider.
A stable Bloch-type SkL generated numerically for this choice of parameters
is shown in Fig. \ref{Fig_SkL}.

\subsection{Thermal dynamics of the spin system}

\label{Sec_dynamics}

We will describe the dynamics of the system by the undamped Landau-Lifshitz
(LL) equation 
\begin{equation}
\hbar\dot{\mathbf{s}}_{i}=\mathbf{s}_{i}\times{\bf H}_{{\rm eff},i},\quad{\bf H}_{{\rm eff},i}\equiv-\frac{\partial\mathcal{H}}{\partial\mathbf{s}_{i}},\label{LL_eq}
\end{equation}
in which the nonlinearity already provides an intrinsic damping due
to the interaction of spin waves of finite amplitude and at nonzero
temperatures. The temperature effects can be taken into account by
adding the coupling to the environment via stochastic fields $\boldsymbol{\zeta}_{i}(t)$
and the corresponding damping \cite{lanlif35}, i. e., by going over
to the Landau-Lifshitz-Langevin (LLL) equation. The LLL equation can
be conveniently solved numerically using the pulsed-noise model \cite{gar17pre},
where one can use high-order integrators with a larger integration
step between the pulses instead of the mainstream second-order Heun
scheme which requires a small integration step. However, there are
objections to using the LLL equation. First, the damped Landau-Lifshitz
equation yields the formulas for the damping of spin waves in terms
of the phenomenological damping constant which differ from the damping
computed microscopically taking into account spin-wave processes (intrinsic
damping). Of course, both channels can contribute to the damping,
however, mixing microscopic and phenomenological terms is methodologically
questionable. Second, at least in dielectrics, the main source of
the environmental damping is due to the fluctuations of the crystal
field driven by the phonons. This creates fluctuating anisotropy rather
than fluctuating field. It was shown that the form of the damping
term in the LL equation, the famous double-vector product, follows
from the symmetry of the Langevin forces \cite{garishpan90}. In the
realistic case of a fluctuating anisotropy, the form of the damping
term becomes more complicated \cite{garishpan90} and inconvenient
because of too many parameters characterizing the crystal field. Third,
if one decides to use a very small phenomenological damping just to
fix the system's temperature, it will become comparable with the effective
damping or antidamping due to the numerical errors, and the results
will be compromised.

Here, we use the concept of the configurational temperature to describe
thermal dynamics of our many-spin system. In Ref. \cite{rugh97prl}
by considering the constant-energy hypersurface of the system of interacting
classical particles it was shown that the temperature can be represented
as a function of their positions and momenta. Strictly speaking, this
function should be averaged over the ergodic trajectory of the system,
but for a large system this averaging can be discarded. This configurational
temperature is not a conserved quantity but fluctuates as the system
moves over its ergodic trajectory, the fluctuations decreasing with
the system size. For classical spins, configurational temperature
was derived in Ref. \cite{nurscho00pre} in a general form. Working
formulas for different spin systems can be obtained from the result
of Ref. \cite{nurscho00pre} or with the help of the Langevin formalism
\cite{madudsemwoo10pre,gar21pre}. For our model without single-site
interactions, the spin temperature is given by
\begin{equation}
T_{S}\equiv\frac{1}{2}\frac{\sum_{i}\left(\mathbf{s}_{i}\times\mathbf{H}_{\mathrm{eff},i}\right)^{2}}{\sum_{i}\mathbf{s}_{i}\cdot\mathbf{H}_{\mathrm{eff},i}},\label{TS}
\end{equation}
where the effective field is defined in Eq. (\ref{LL_eq}) and the
Boltzmann constant is discarded. If $T\rightarrow0$, spins align
with their effective fields, thus $T_{S}\rightarrow0$. If $T\rightarrow\infty$,
spins decorrelate from their effective fields, thus the denominator
goes to zero and $T_{S}\rightarrow\infty$. Spin temperature $T_{S}$
of a large system in a state obtained by Monte Carlo at a set temperature
$T$ is close to $T$ with small fluctuations inherent in both methods.
The formula for $T_{S}$ with single-site interactions (uniaxial coherent
or random anisotropy) can be found in Ref. \cite{gar21pre}, while
the generalization for vector spins with an arbitrary number of components
was done in Ref. \cite{kanloisch05epjb}

To formulate the thermal dynamics of a spin system, we use the spin
temperature formula together with the energy-correction procedure
proposed in Ref. \cite{gar21pre}. to beat the accumulation of numerical
errors in long dynamics computations. To correct the energy per spin
by the amount
\begin{equation}
\delta E=E_{\mathrm{target}}-E,\label{deltaE}
\end{equation}
where $E$ is the actual energy of the system subject to the error
accumulation and $E_{\mathrm{target}}$ is its correct value (for
instance, computed via the formula for the energy absorbed by the
system irradiated by microwaves \cite{garchu22prb}), one is rotating
the spins as 
\begin{equation}
\delta\mathbf{s}_{i}=\xi\mathbf{s}_{i}\times\left(\mathbf{s}_{i}\times\mathbf{H}_{\mathrm{eff},i}\right),\qquad\xi=\frac{\delta E}{(1/N)\sum_{i}\left(\mathbf{s}_{i}\times\mathbf{H}_{\mathrm{eff},i}\right)^{2}}\label{delta_s_i}
\end{equation}
with the subsequent normalization of the new spins $\mathbf{s}_{i}+\delta\mathbf{s}_{i}$.
This corrects the energy to the leading order in $\delta E$ which
should be kept small by performing energy correction frequently enough,
depending on the accuracy of the numerical integrator of the Landau-Lifshitz
equation. As said above, one way of correcting the system energy is
adjusting it to the absorbed MW energy per spin
\begin{equation}
E_{{\rm abs}}(t)=-\frac{1}{N}\int_{0}^{t}dt'\,\dot{\mathbf{h}}(t')\cdot\sum_{i}\mathbf{s}_{i}(t')\label{E_abs_def}
\end{equation}
requiring $E(t)-E(0)=E_{{\rm abs}}(t)$. As the absorbed energy, once
accounted for, does not change, this keeps also the system energy
correct over long times. 

Here we modify this approach and set 
\begin{equation}
\delta E=-\eta\left(T_{S}-T\right),\label{Delta_E_Thermal}
\end{equation}
where $T_{S}$ is the spin temperature, $T$ is the environmental
temperature, and $\eta$ is number proportional to the environmental
coupling constant. If the energy-correction procedure is repeated
at small time intervals $\delta t$, then, dividing this formula by
$\delta t$, one obtains the continuous form of the energy-correction
procedure:
\begin{equation}
\dot{E}=-\kappa\left(T_{S}-T\right),\label{Edot_eq}
\end{equation}
 where $\dot{E}\equiv\delta E/\delta t$ and $\kappa\equiv\eta/\delta t$.
This is nothing else than the standard macroscopic equation describing
the energy exchange between the system and the bath. If, for instance,
$T_{S}>T$, the energy $E$ is decreased which, in turn, leads to
the decrease of $T_{S}$, as these quantities are implicitly related
and both increase or both decrease. With this approach, one can stabilize
$T_{S}$ close to $T$ and the computation can run infinitely long
as there is no problem with the error accumulation.

One also can formulate the continuous heat exchange with the bath
in terms of lattice spins. From Eq. (\ref{delta_s_i}) one obtains
the relaxation law for the spins 
\begin{equation}
\dot{\mathbf{s}}_{i}=-\kappa\frac{T_{S}-T}{(1/N)\sum_{i}\left(\mathbf{s}_{i}\times\mathbf{H}_{\mathrm{eff},i}\right)^{2}}\mathbf{s}_{i}\times\left(\mathbf{s}_{i}\times\mathbf{H}_{\mathrm{eff},i}\right).\label{s_i_relax}
\end{equation}
Adding this to the undamped LL equation, one obtains
\begin{equation}
\hbar\dot{\mathbf{s}}_{i}=\mathbf{s}_{i}\times{\bf H}_{{\rm eff},i}-\alpha_{\mathrm{eff}}\mathbf{s}_{i}\times\left(\mathbf{s}_{i}\times\mathbf{H}_{\mathrm{eff},i}\right),\label{LLG_eq}
\end{equation}
where the effective damping constant is given by
\begin{equation}
\alpha_{\mathrm{eff}}=\frac{\hbar\kappa\left(T_{S}-T\right)}{(1/N)\sum_{i}\left(\mathbf{s}_{i}\times\mathbf{H}_{\mathrm{eff},i}\right)^{2}}.\label{alpha_eff}
\end{equation}
For many-spin systems, the damped Landau-Lifshits equation above must
be more convenient than the Landau-Lifshitz-Langevin equation. The
computation of $T_{S}$ in its numerical solution will slow it down
but not critically as it takes $\sim N$ operations, same as computing
the right-hand side of the LL equation. The relaxation term should
not contribute much into the spin-wave damping because at equilibrium
$T_{S}$ is very close to $T$ and this term is small. In this work,
we will use the discrete energy corrections using Eq. (\ref{Delta_E_Thermal})
which causes no slow-down at all.

The expression for the absorbed power of the MW field of the frequency
$\omega$ linearly polarized in the $\alpha$-direction ($\alpha=x,y,z$)
is provided by the fluctuation-dissipation theorem (FDT) (see, e.g.,
Ref. \cite{garchu22prb}): 
\begin{equation}
P_{\alpha}(\omega)=\frac{N\omega^{2}h_{0}^{2}}{2k_{B}T}\mathrm{Re}\left[\int_{0}^{\infty}dte^{i\omega t}\left\langle m_{\alpha}(t_{0})m_{\alpha}(t_{0}+t)\right\rangle _{t_{0}}\right].\label{FDT}
\end{equation}
Here $N$ is the total number of spins,
\begin{equation}
{\bf m}(t)=\frac{1}{N}\sum_{i}{\bf s}_{i}\label{m_def}
\end{equation}
 is the magnetization (spin polarization) of the system, $h_{0}$
is the amplitude of the MW field, and $T$ is the temperature. The
averaging in Eq. (\ref{FDT}) is done over $t_{0}$, the starting
point of the $t$-interval. That is, we average over the time rather
than over a thermodynamic ensemble which is computationally more efficient.
To obtain an accurate absorption spectrum of the system with noise
averaged out, one needs a rather long computation. An advantage of
this method is that it allows to obtain results for all frequencies
froma single computation run (see e.g., an application of this method
to MW absorption in random-anisotropy magnets in Ref. \cite{garchu22prb}). 

\subsection{Skyrmion lattice}

\label{Sec_SkL}

The degree of the crystalline order in a SkL is measured by translational
and orientational order parameters \cite{HN-PRL1978,NH-PRB1979}.
The translational order shows how perfect the SkL is with respect
to the translations by a multiple of the lattice spacing $a_{S}$
which is the distance between neighboring skyrmions in a perfect lattice.
It is related to the structure factor, 
\begin{equation}
S({\bf q})=\sum_{i}e{}^{i{\bf q}\cdot{\bf r}_{i}},
\end{equation}
where ${\bf r}_{i}$ is the position of the $i$-th skyrmion in the
SkL. For a perfect hexagonal lattice $S({\bf q})$ has sharp maxima
at ${\bf q}$ equal one of the reciprocal lattice vectors or their
linear combinations. For the SkL with horizontally oriented hexagons,
such as in Fig. \ref{Fig_SkL}, the reciprocal vectors are 
\begin{equation}
{\bf q}_{1}=(0,1)q,\quad{\bf q}_{2}=\frac{(\sqrt{3},-1)}{2}q,\quad{\bf q}_{3}=\frac{(-\sqrt{3},-1)}{2}q,\label{reciprocal}
\end{equation}
where $q=4\pi/(\sqrt{3}a_{S})$ The translational order parameter
can be defined as 
\begin{equation}
O_{tr}=\frac{1}{3N_{S}}\sum_{\nu=1}^{3}\left|S({\bf q}_{\nu})\right|,\label{Otr_def}
\end{equation}
where $N_{S}$ is the total number of skyrmions in the SkL and ${\bf q}_{v}$
are the three reciprocal-lattice vectors given by Eq.\ (\ref{reciprocal}).
For a perfect lattice, $O_{tr}=1$. It becomes smaller than $1$ in
the presence of thermal fluctuations and/or lattice defects, and it
is completely destroyed as the skyrmion lattice melts.

The quantity describing the orientation of a hexagon formed by the
nearest neighbors of the $i$-th skyrmion is 
\begin{equation}
\Psi_{i}=\frac{1}{6}\sum_{j=1}^{6}\exp\left(6i\theta_{ij}\right),
\end{equation}
where the summation is over the six nearest neighbors denoted by $j$,
and $\theta_{ij}$ is the angle that the $ij$-th bond makes with
a fixed direction in the lattice which we choose to be in the positive
$x$-direction, that is, $(1,0)$. For a perfect SkL, $|\Psi_{i}|=1$.
Thermal disorder and/or dislocations make $|\Psi_{i}|$ smaller than
$1$. In the presence of disorder, the quality of hexagons can be
evaluated by computing the quantity 
\begin{equation}
V_{6}=\sqrt{\frac{1}{N_{S}}\sum_{i}|\Psi_{i}|^{2}}.
\end{equation}
The orientational order parameter, describing the common orientation
of hexagons in a SkL is 
\begin{equation}
O_{6}=\frac{1}{N_{S}}\sum_{i}\Psi_{i}.
\end{equation}
If $V_{6}$ and $O_{6}$ are computed by Monte Carlo or another process,
they can be averaged over this process after equilibration, to improve
their accuracy. Melting of the SkL to the skyrmion liquid destroys
$O_{6}$ while $V_{6}$ decreases but remains finite. In the high-temperature
limit the angles $\theta_{ij}$ become random and $V_{6}\rightarrow\sqrt{1/6}$.
To account for the degree of disorder in the mutual orientations of
hexagons irrespective of their quality, it is also convenient to compute
$O_{6}/V_{6}$.

The excitation spectrum of the SkL contains three uniform modes \cite{Mochizuki-PRL2012}.
In the breathing mode, the $xy$ components of the spins on the skyrmions'
slopes intermittently rotate clockwise and counterclockwise which
results in the oscillations of the skyrmion size due to the DMI and
the ensuing oscillations of $m_{z}$. Two other modes correspond to
the precession of the spins around the $z$-axis in the regions of
skyrmions cores (spins up) and in the regions between skyrmions (spins
down) \cite{Mochizuki-PRL2012,Satywali-NatCom2021}. In these modes,
spins precess in different directions \cite{Mochizuki-PRL2012}. Within
the model with DMI, as in Ref. \cite{Mochizuki-PRL2012} and here,
the low-frequency (LF) mode is stronger and localized in the regions
between the skyrmions. It corresponds to the bulk mode in the model
with a single skyrmion. Another, the high-frequency (HF) mode, is
localized on skyrmions. In Ref. \cite{Satywali-NatCom2021} in a particular
compound it was observed that, vice versa, the LF mode is localized
on skyrmions and HF mode is localized in the regions between them.
Thus calling the modes LF and HF may be misleading and it would be
better to use the terms ``core more'' and ``bulk mode''.

\section{Numerical method}

\label{Sec_numerical_method}

In numerical calculations, we set $J=a=k_{B}=\hbar=g\mu_{B}=1$, as
usual. These constant are also dropped from the text below, except
for $J$. The actual time is related to the dimensionless computing
time $t$ we show as $t_{\mathrm{actual}}=t\hbar/J$. Whereas $t$
may be very long, up to $10^{6}$ in our computations, the actual
time in seconds, passing in experiments, is typically by a factor
of $10^{12}$ shorter.

To create a skyrmion lattice, we start with a lattice of arbitrary-shape
spin bubbles in proper SkL positions in the system with periodic boundary
conditions (pbc). The sizes $N_{x}$ and $N_{y}$ of the system (in
the units of the lattice spacing $a$) are chosen so that the system
shape is close to a square and the system accommodates the SkL without
distortions. As the main set of parameters we use $A/J=0.2$, $H/J=-0.025$
and the period of the skyrmion lattice $a_{S}=38.5a$. The details
and illustrations can be found in Ref. \cite{DG-EC-JMMM2024}.

\begin{figure}
\begin{centering}
\includegraphics[width=8cm]{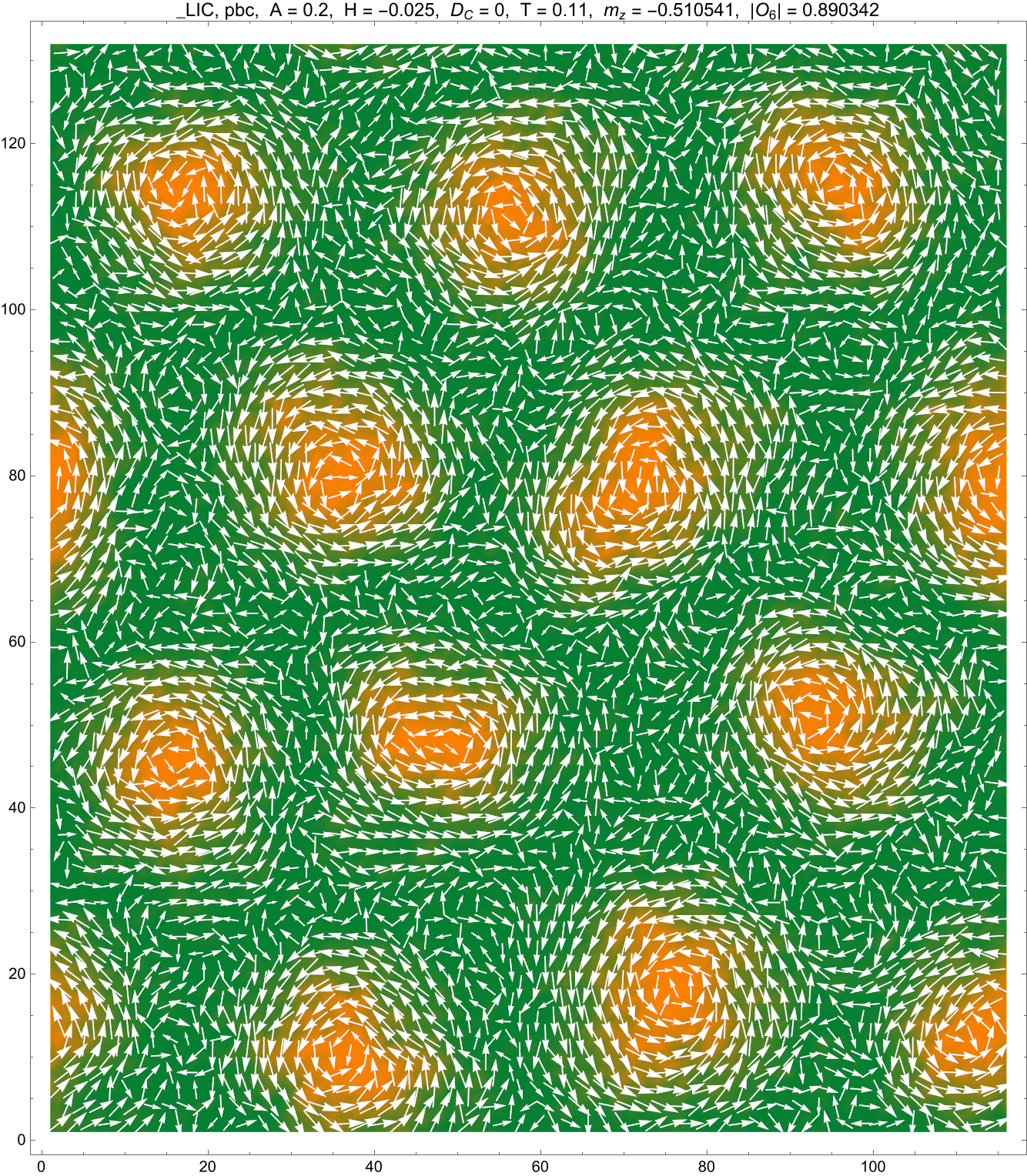}
\par\end{centering}
\caption{Thermalized SkL at $T/J=0.11$, cf. Fig. \ref{Fig_SkL}.}

\label{Fig_SkL_thermalized}
\end{figure}

Then we run the energy-minimization routine at $T=0$ which consists
in sequential rotation of spins $\mathbf{s}_{i}$ toward their effective
field $\mathbf{H}_{\mathrm{eff},i}=-\partial\mathcal{H}/\partial\mathbf{s}_{i}$
(field alignment) with the probability $\alpha$ and rotating the
spins by 180$^{\circ}$ around $\mathbf{H}_{\mathrm{eff},i}$ (the
energy-conserving overrelaxation) with the probability $1-\alpha$.
We used $\alpha=0.01-0.03$ which allows an efficient exploring of
the phase space of the system by overrelaxation and ensures a much
faster convergence than the field alignment alone ($\alpha=1)$. At
$T>0$, we run Metropolis Monte Carlo, also combined with overrelaxation
with the same $\alpha$, after energy minimization or without preceding
energy minimization. These preparatory stages of the computation cost
only a small fraction of the total computing time. An example of a
thermalized SkL is shown in Fig. \ref{Fig_SkL_thermalized} for $T/J=0.11$
(just below the SkL melting point) to show the effect of the thermal
disordering clearly. In most subsequent computation much lower temperatures
are used, mainly $T/J=0.01$.

In the FDT part of the work which does not include the microwave field,
we solve Eq. (\ref{LL_eq}) numerically starting from a prepared state
over a long time interval using Butcher's 5th-order Runge-Kutta integrator
which makes 6 evaluations of the right-hand side of the equation of
motion per integration step. Since this method is rather precise,
we use a large integration step $\Delta t=0.1J$. Computations are
done in time intervals comprising, typically, 70 integration steps.
After completing a time interval, the values of the magnetization
$\mathbf{m}$, Eq. (\ref{m_def}), are recorded and the energy correction
procedure according to Eq. (\ref{Delta_E_Thermal}) is performed.
The latter takes a very short time. Due to the energy correction,
the temperature and the energy of the system are kept constant up
to small fluctuations. From time to time, we interrupt the main computation
to compute the power absorption spectra $P_{\alpha}(\omega)/h_{0}^{2}$
and monitor the appearance of absorption peaks and decrease of the
noise. The average over $t_{0}$ in the time correlation functions
(CFt) in Eq. (\ref{FDT}) is done with the help of the convolution
of discrete time series which is internally using the Fast Fourier
Transform and thus is fast. Then we compute the Fourier transform
of the CFt to find $P_{\alpha}(\omega)/h_{0}^{2}$. The process is
terminated when the absorption peaks become smooth enough (see Sec.
\ref{Sec_FDT}). The breathing mode is seen in $P_{z}(\omega)/h_{0}^{2}$
whereas $P_{x}(\omega)/h_{0}^{2}$ contains two peaks corresponding
to the LF and HF modes. (As $P_{x}=P_{y}$, the result can be symmetrized).
The positions of the peaks yields the frequencies of the three uniform
excitation modes of the SkL.

One can also find the temperature dependence of the peak widths $\Gamma$
which is easier to extract from the time-dependent correlation functions.
The $\left\langle m_{z}m_{z}\right\rangle $ CFt contains only one
mode and can be approximated by $e^{-\Gamma_{z}t}\cos\left(\omega_{z}t\right)$.
The frequency $\omega_{z}$ can be extracted from the time differences
between the extrema of this function. The damping $\Gamma_{z}$ can
be extracted from the envelope. This requires a rather long computation
before CFt stabilizes and fluctuations decrease. Extracting $\Gamma_{z}$
is more difficult and the result has a lower accuracy than that for
$\omega_{z}$.

As there are two different modes contributing to the $\left\langle m_{x}m_{x}\right\rangle $
CFt, it has, in general a more complicated form which depends on the
phases of the two modes in a given state which are random. This difficulty
can be overcome by considering the absorption of circularly-polarized
microwaves with different rotations, each containing only one absorption
peak. This requires an appropriate modification of Eq. (\ref{FDT}).
On the other hand, in most situations the regions near the skyrmions'
tops are small in comparison with the regions between the skyrmions
which makes the HF mode weak in comparison to the LF mode. Only for
very small values of $H$ when skyrmions become broad and tightly
squeezed, these two modes become of comparable strengths. Thus, in
extracting the lines' widths for $H/J=-0.025$, we just ignore the
weak HF mode and find the damping of the LF mode by the same method
as above. The damping of the weak HF mode should be very difficult,
if not impossible, to find.

In the experiments with pumping the SkL with microwaves, we explore
the two cases: (i) the system isolated from the environment and (ii)
the system coupled to the environment. As the systems under investigation
are thin films on substrates, the physically relevant case is the
second one. For the isolated system, we correct the system's energy
to the absorbed MW energy, Eq. (\ref{E_abs_def}). For the coupled
system, we use the energy correction given by the thermal-exchange
firmula, Eq. (\ref{Delta_E_Thermal}). The energy-correction procedure
must be done only at the moments of time when the MW field $h(t)=h_{0}\sin\left(\omega t\right)$
is zero. 

To compute the parameters of the SkL under the influence of microwaves,
one fist has to locate skyrmions. We define skyrmions as regions ``above
the sea level'' which is set to 0.5. That is, we are looking for
``islands'' with $s_{z}>0.5$. To enumerate all of them, we used
a variant of the single-pass connected-component labeling algorithm
described in more detail in Ref. \cite{DG-EC-JMMM2024}. If a spin
with $s_{z}>0.5$ is found, it is added to a new island, then all
its neighbors are checked and added to the same island if they have
$s_{z}>0.5$. This ends when there are no more neighbors above the
sea level. If the number of spins within the island is greater or
equal to the minimal island size, a skyrmion is identified. The limitation
on the skyrmion size is important at elevated temperatures when spins
are substantially disordered and there can be even isolated spins
pointing up which should not be counted as skyrmions. After all skyrmions
are identified, positions of their centers $\mathbf{R}_{i}$ can be
found with the help of the skyrmion-locator formula \cite{DG-EC-JMMM2024}
\begin{equation}
\mathbf{R}_{i}=\sum_{j\in i}\mathbf{r}_{j}s_{z,j}^{2}/\sum_{j\in i}s_{z,j}^{2},\label{Skyrmion_locator}
\end{equation}
where $j\in i$ are all lattice sites that belong to the skyrmion
$i$. Here the weight factor $s_{z,j}^{2}$ favors the sites closer
to the skyrmion's top.

\section{Numerical results}

\label{Sec_Numerical_results}

\subsection{Resonance frequencies of SkL's uniform modes by the FDT}

\label{Sec_FDT}

\begin{figure}
\begin{centering}
\includegraphics[width=8cm]{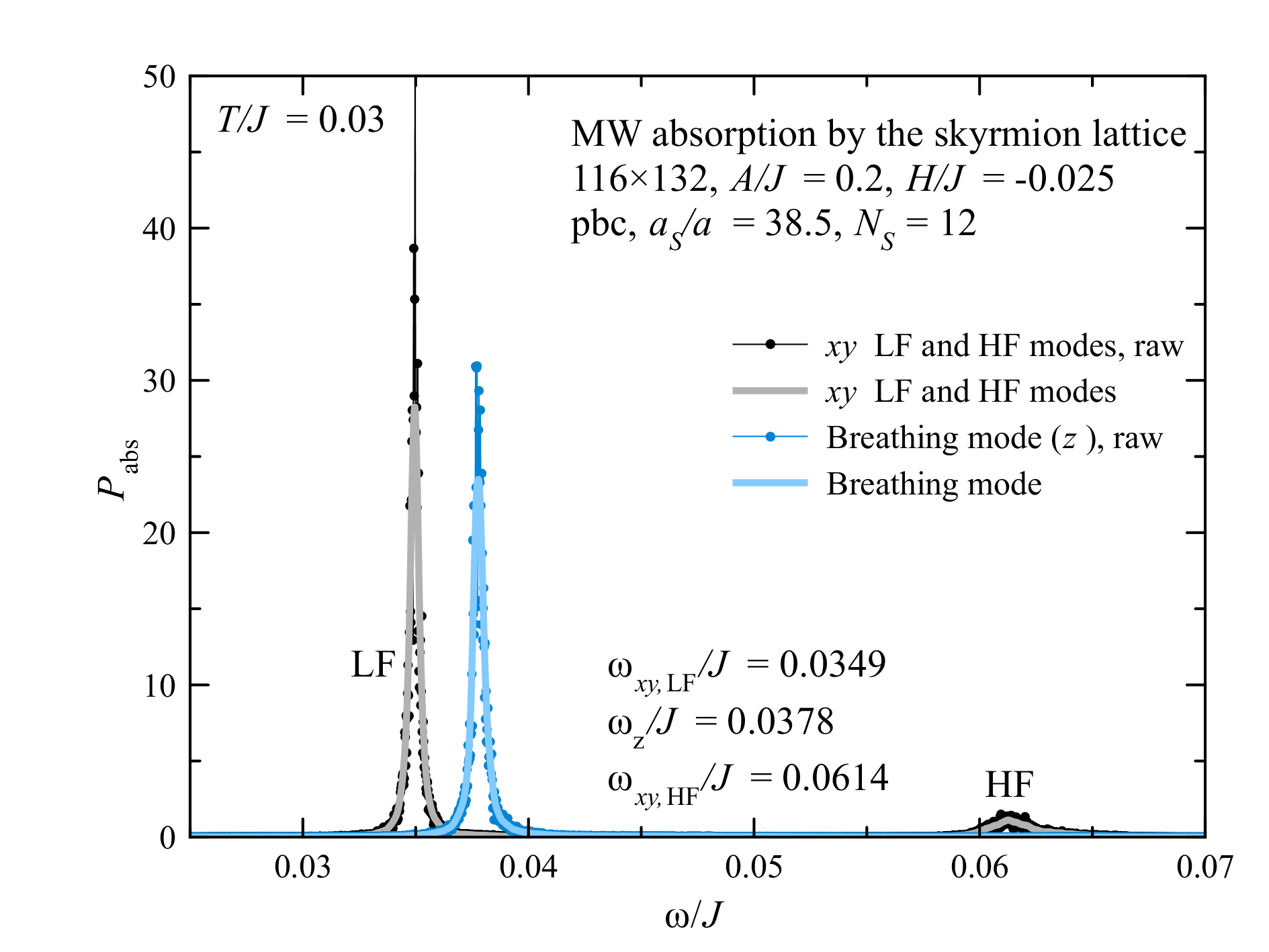} 
\par\end{centering}
\centering{}\includegraphics[width=8cm]{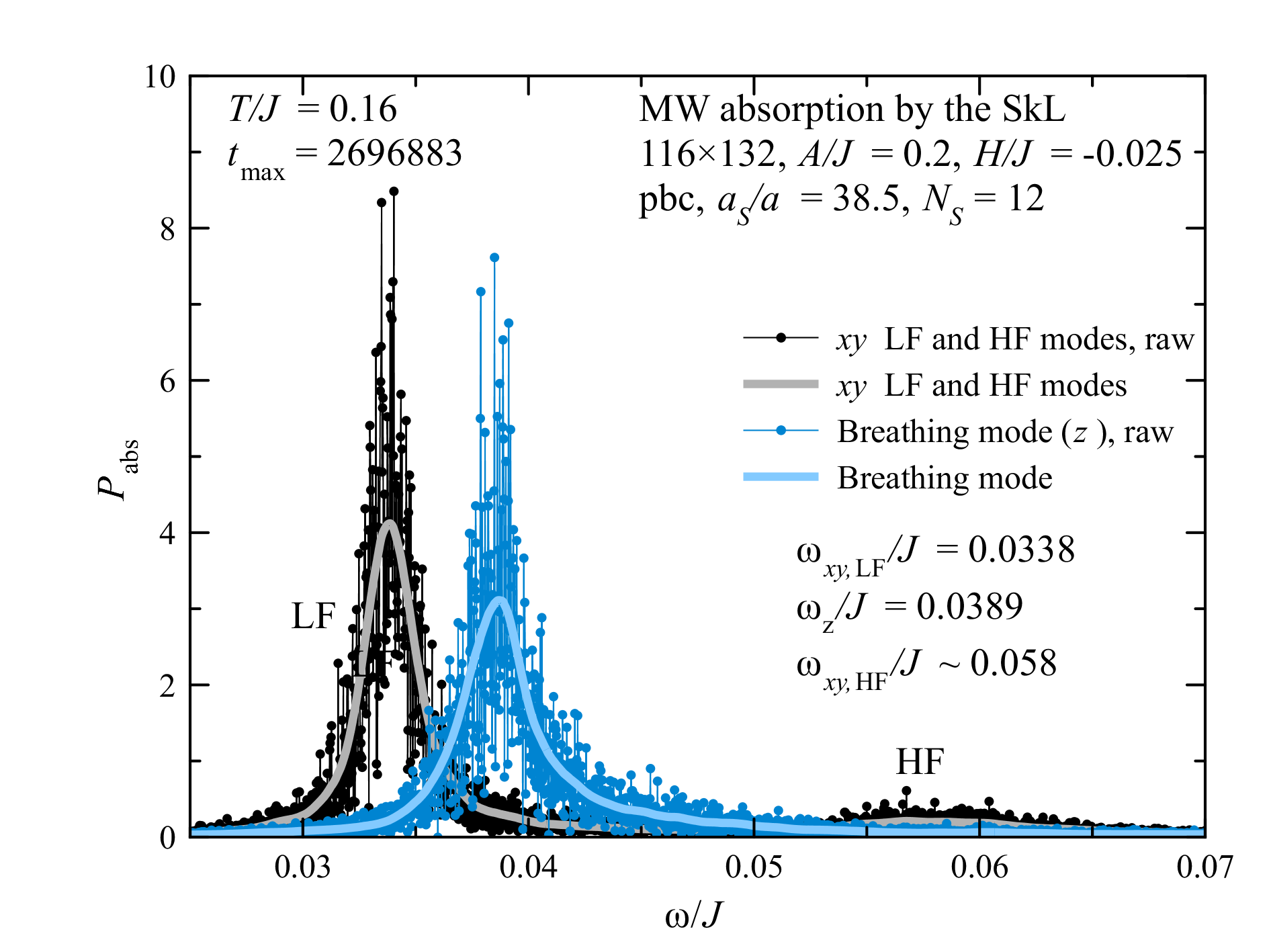}\caption{Power-absorption spectra of the skyrmion lattice of $116\times132$
spins containing 12 skyrmions for our main set of parameters. Upper:
$T/J=0.03$; Lower: $T/J=0.16$. Note the temperature dependence of
the modes' frequencies and damping constants.}
\label{Fig_Pabs} 
\end{figure}

The power-absorption spectrum of the skyrmion lattice containing 12
skyrmions for our main parameters set is shown at different temperatures
in Fig. \ref{Fig_Pabs} where the results for $P_{z}(\omega)$ (breathing
mode) and $\left(P_{x}(\omega)+P_{y}(\omega)\right)/2$ (LF and HF
modes) are combined. There are both raw and smoothed data. At our
low temperature, $T/J=0.03$, the peaks in Fig. \ref{Fig_Pabs} (upper)
are narrow with not much scatter. As said above, for this value of
$H$ the HF peak is weak. It is noteworthy that even here the intrinsic
damping is strong enough to render the peaks an appreciable width
so that no phenomenlogical damping is needed to visualize them. Also
interesting is that here the frequency of the breathing mode is higher
than that of the LF mode. For an isolated skyrmion, to the contrary,
the frequency of the breathing mode \cite{DG-RJ-EC-PRB2020} $\omega_{\mathrm{breathing}}\cong0.8|H|$
is below that of the bulk mode, $\omega_{\mathrm{bulk}}=|H|$, while
the bulk mode is transforming into the LF mode if skyrmions are added
to the system. This difference must be due to the interaction between
skyrmions in the SkL. 

At elevated temperatures, as for $T/J=0.16$ (above the SkL melting
point) in Fig. \ref{Fig_Pabs} (lower), absorption peaks become much
broader and the noise strongly increases. To reduce this noise, a
very long computation up to the time $t_{\max}=2.7\times10^{6}$ was
needed. Note that the frequencies of the peaks changed in comparison
to the low-temperature case: the frequencies of the LF and HF modes
decreased but that of the breathing mode increased. Usually, excitation
modes soften with increasing the temperature, this the behavior of
the breathing mode is unusual.

\begin{figure}
\begin{centering}
\includegraphics[width=8cm]{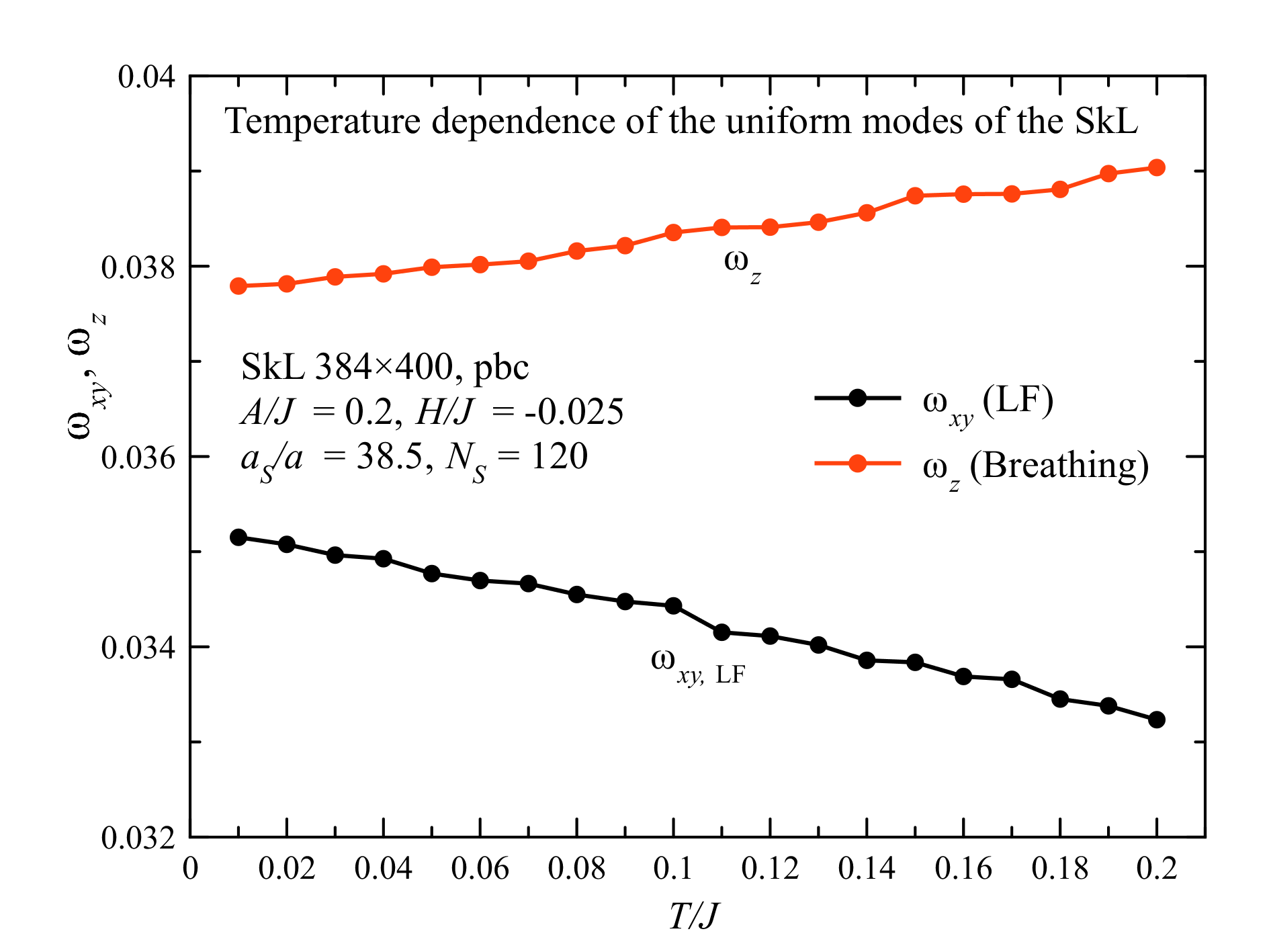} 
\par\end{centering}
\centering{}\includegraphics[width=8cm]{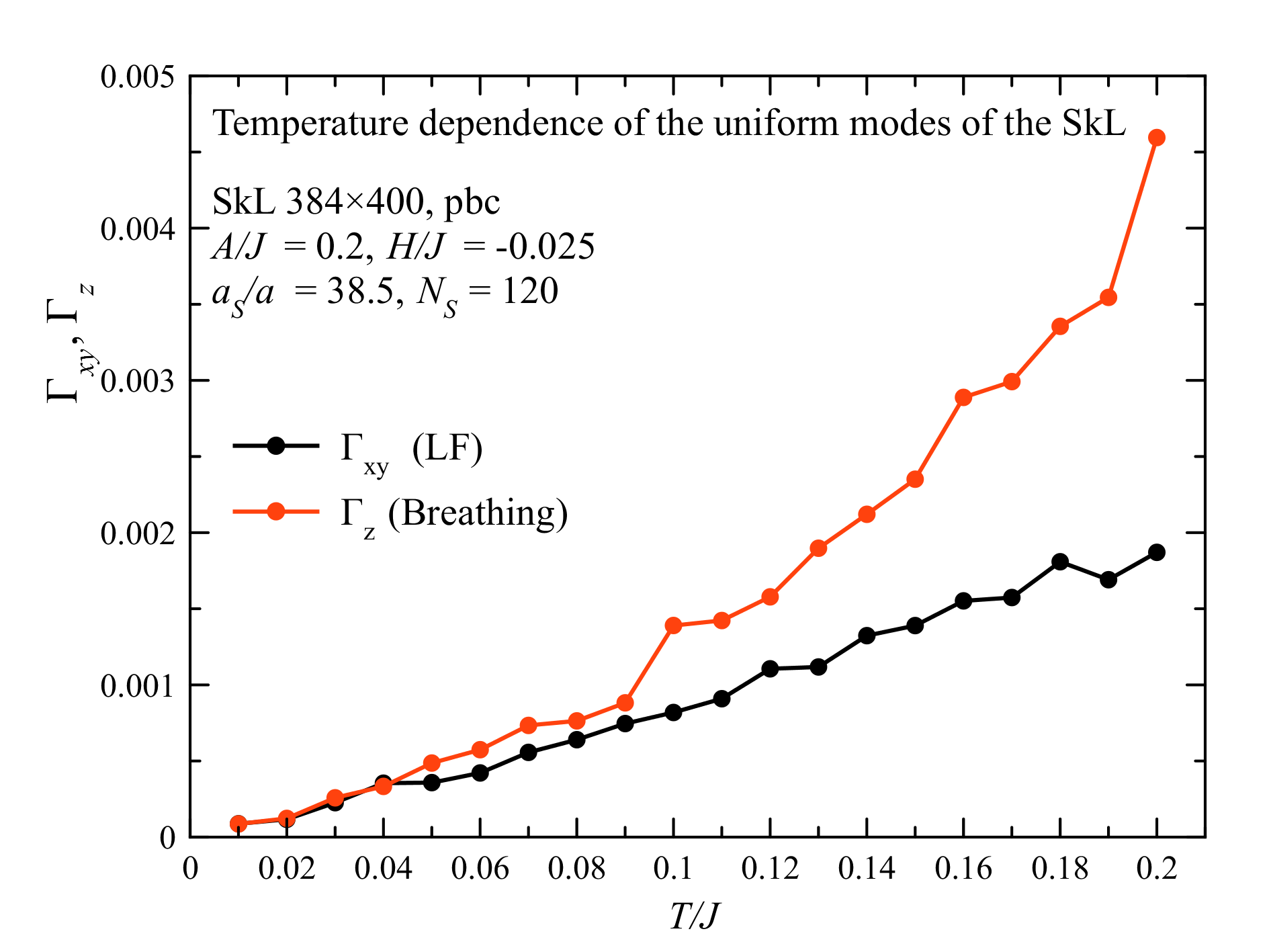}\caption{The temperature dependence of the frequency and intrinsic damping
of the breathing and LF modes for the system of $384\times400$ spins
containing 120 skyrmions. Upper: frequencies; Lower: damping.}
\label{Fig_Modes_frequencies_and_damping} 
\end{figure}

Fig. \ref{Fig_Modes_frequencies_and_damping} shows the temperature
dependence of the frequency and damping of the breathing and LF modes
extracted from the time correlation functions as explained in Sec.
\ref{Sec_numerical_method}. Here we use a bigger system of $384\times400$
spins containing 120 skyrmions. In Fig. \ref{Fig_Modes_frequencies_and_damping}
(upper) one can see that the frequencies depend on $T$ linearly,
whereas that of the breathing mode unusually increases. It is noteworthy
that there is no effect of the SkL melting at $T_{m}/J\simeq0.12$
on the the frequencies of the modes. The damping constants of both
modes increase with the temperature, as Fig. \ref{Fig_Modes_frequencies_and_damping}
(lower) shows. Whereas for the LF mode the damping is linear in $T$,
it increases stronger for the breathing mode.

\begin{figure}
\begin{centering}
\includegraphics[width=8cm]{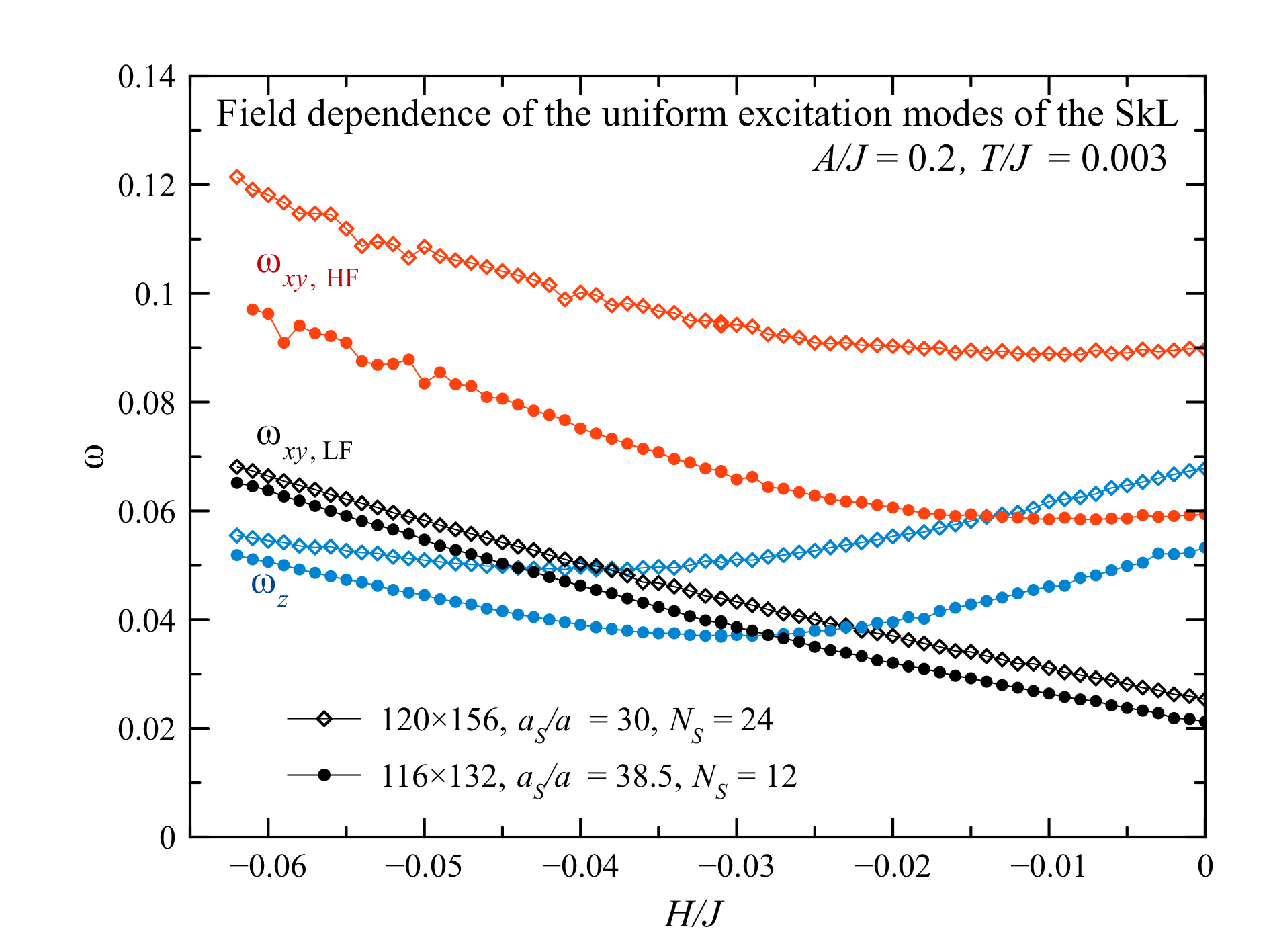}
\par\end{centering}
\caption{Field dependence of the SkL modes' frequencies at nearly zero temperature
for our main SkL and an SkL with a larger density of skyrmions. In
the latter, all three modes stiffen.}

\label{Fig_omega_vs_H}
\end{figure}

Fig. \ref{Fig_omega_vs_H} shows the field dependence of the SkL modes'
frequencies obtained by finding the positions of the peaks in the
absorbed power spectrum, Fig. \ref{Fig_Pabs} at a very low temperature
of $T/J=0.003$. The results were obtained for our small system of
$116\times132$ spins with 12 skyrmions as well as for a larger system
of 24 skyrmions with a higher concentration of skyrmions because of
a smaller SkL period of $a_{S}=30a$. Whereas the system size does
not affect the characteristics of the modes much, with the increasing
of the density of skyrmions all modes stiffen, as expected. As can
be seen, the field dependence of the breathing mode differs from that
of the LF and HF modes so that there is its crossing with the LF mode.
There is no hybridization of the modes near the crossing points (which
was reported in Refs. \cite{Aqeel-PRL2021,Lee-JPhys2022}) within
our model but additional interactions may cause hybridization. On
the left side of this plot, for stronger fields, skyrmions become
small and tend to behave as isolated skyrmions because of the weakened
skyrmion-skyrmion interaction \cite{capgarchu20jpcm}. In this regime,
$\omega_{z}<\omega_{xy,\mathrm{LF}}$, as for a single skyrmion \cite{DG-RJ-EC-PRB2020}.

\subsection{Interaction of a skyrmion crystal with microwaves}

\label{Sec_MW}

\begin{figure}
\begin{centering}
\includegraphics[width=8cm]{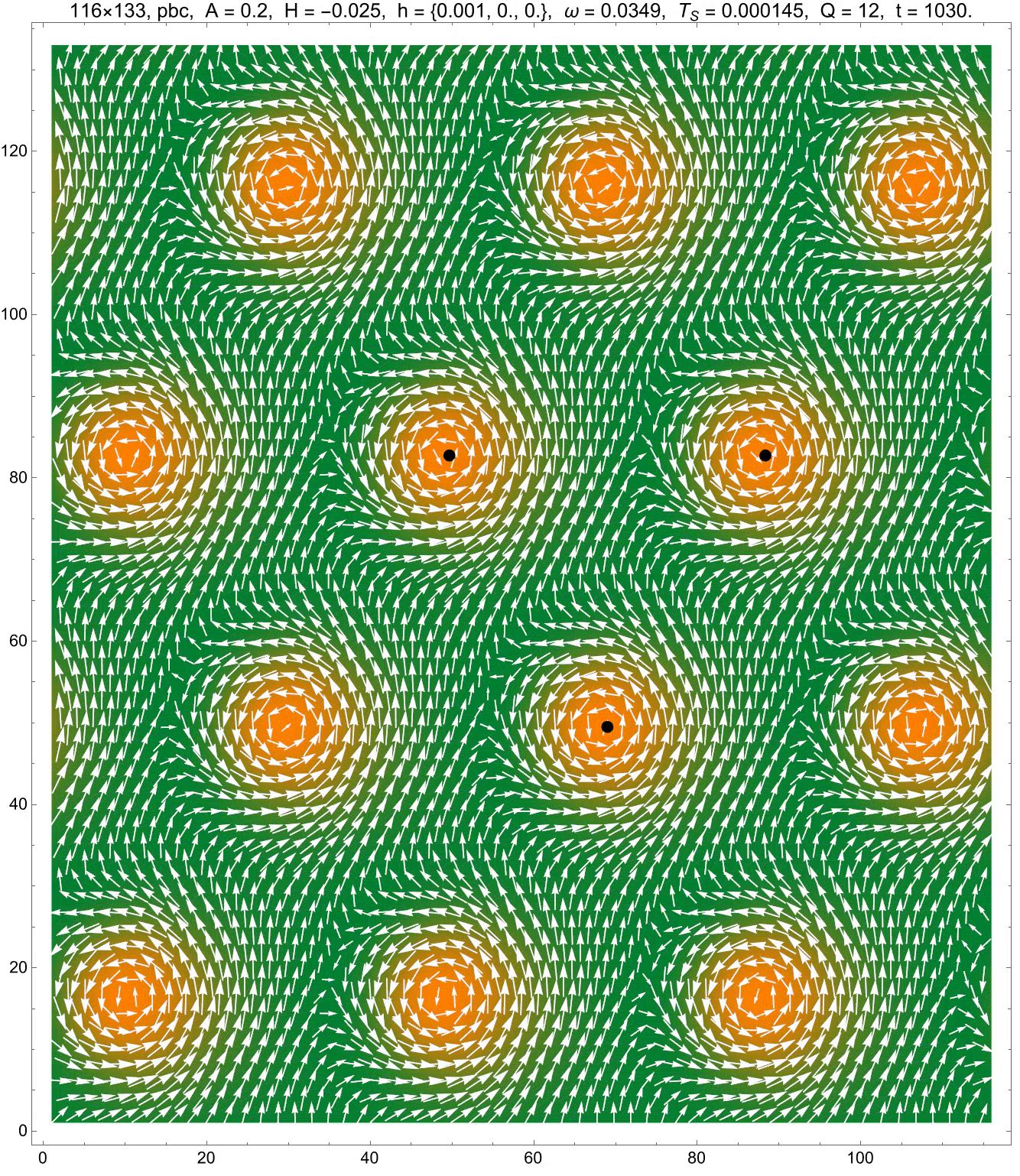}
\par\end{centering}
\begin{centering}
\includegraphics[width=8cm]{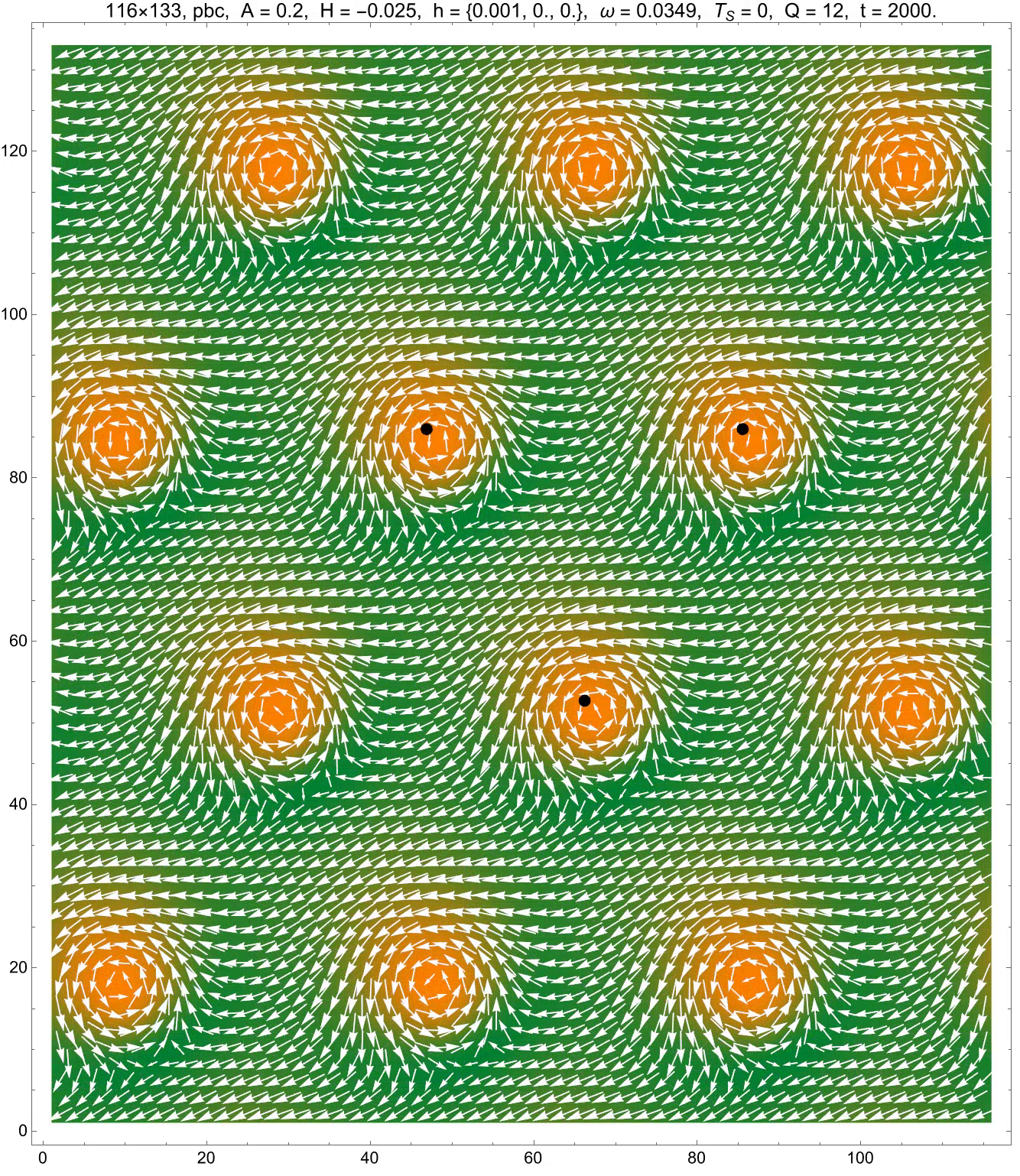}
\par\end{centering}
\caption{Early stages of the resonant pumping of the LF mode: $t=1030$ and
2000. The centers of three skyrmions found by the locator formula,
Eq. (\ref{Skyrmion_locator}), are shown by black dots. Here the spins
in the green regions between skyrmions are precessing in-phase while
skyrmions' centers are rotating around their initial positions. The
amplitude of this motion increases and decreases. As skyrmions becomes
deformed and lose their angular symmetry, their centers found by the
locator formula do not exactly coincide with their tops.}

\label{Fig_LF_pumping_early}
\end{figure}
\begin{figure}
\begin{centering}
\includegraphics[width=8cm]{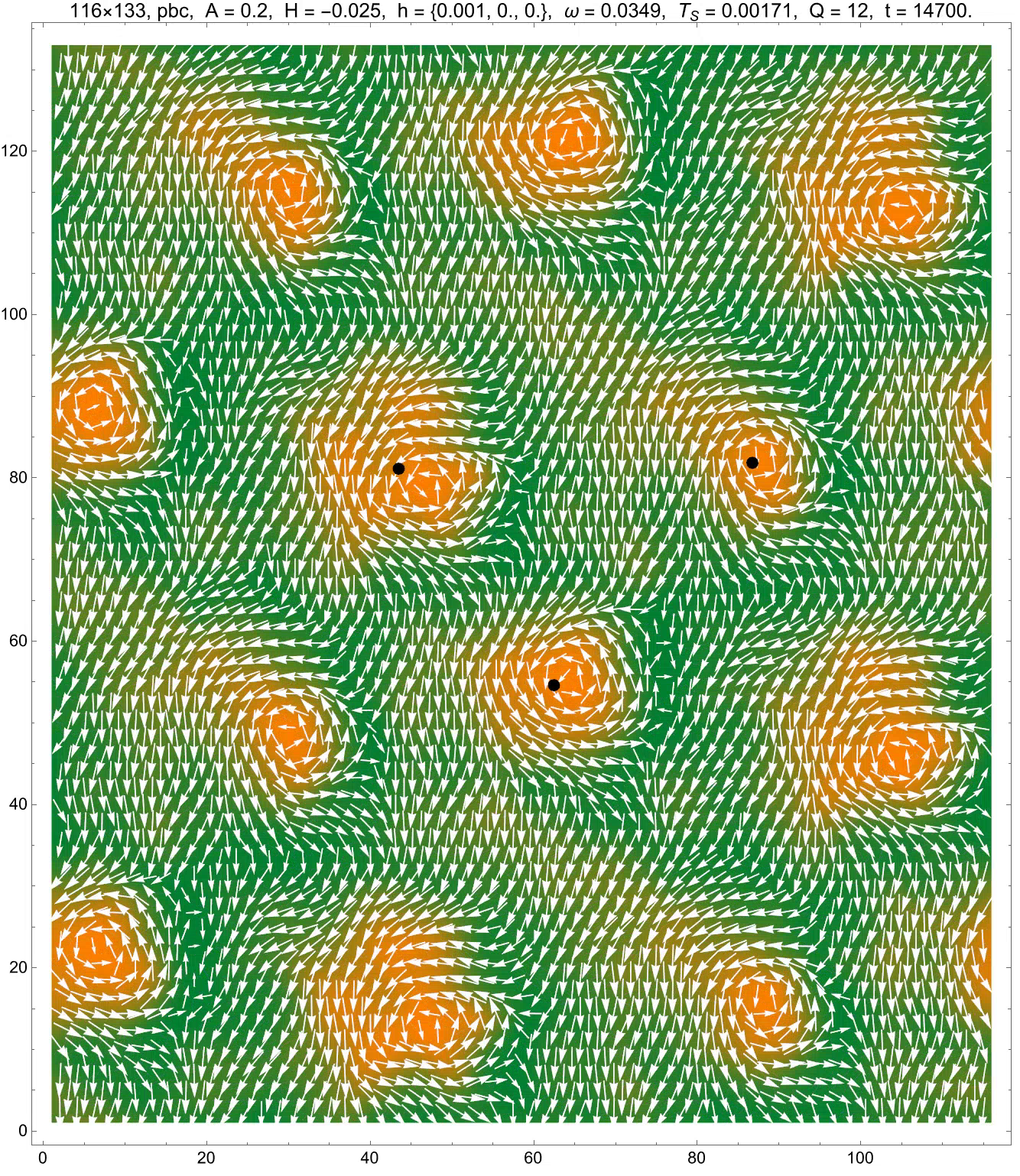}
\par\end{centering}
\begin{centering}
\includegraphics[width=8cm]{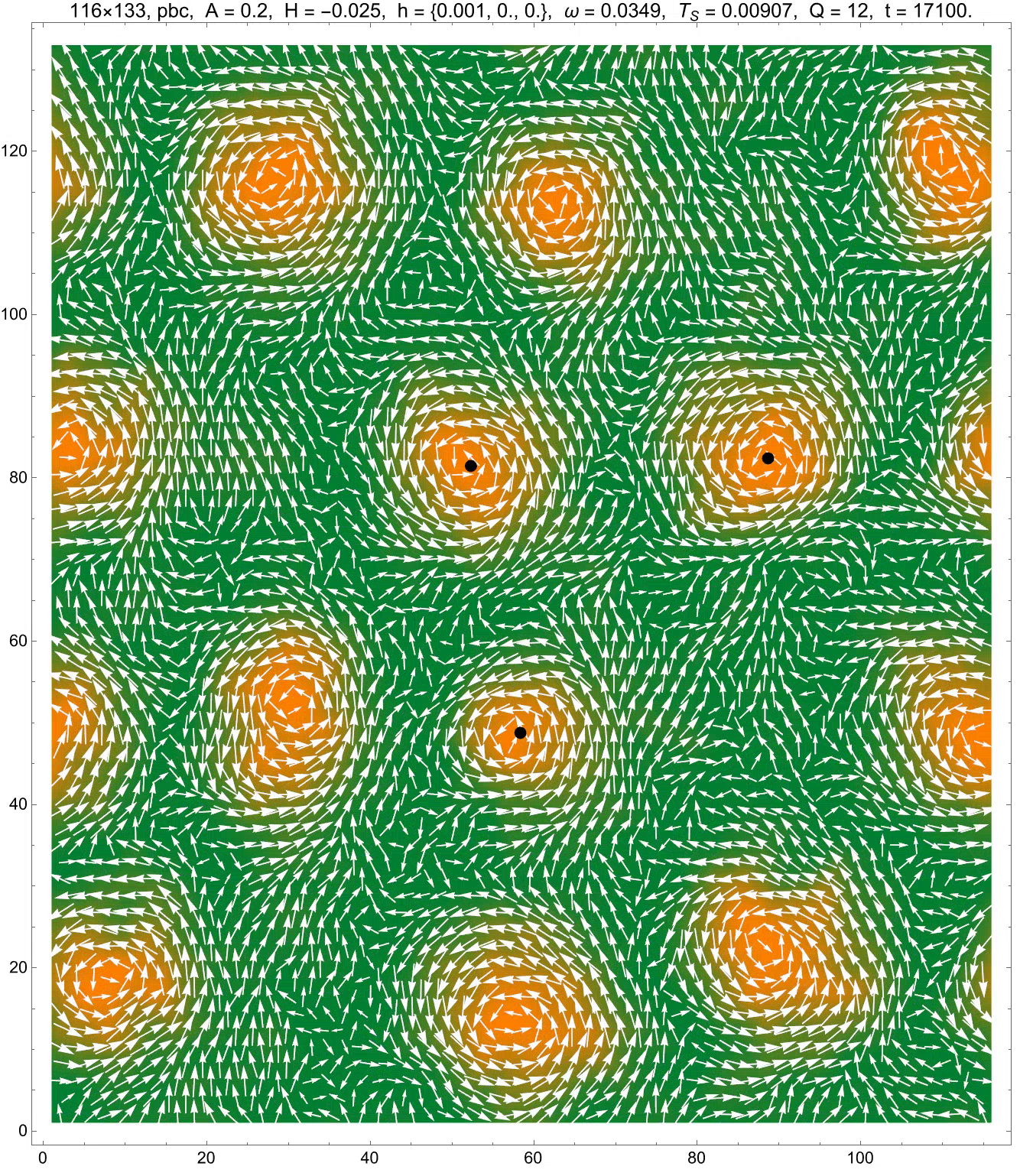}
\par\end{centering}
\caption{Advanced stages of the resonant pumping of the LF mode: $t=14700$
and 17100. Upper panel: precession of spins in the regions between
skyrmions partially lost coherence and skyrmions are strongly deformed
with \textquotedblleft tails\textquotedblright{} in the upper-left
direction at $t=14700$. Lower panel: SkL is destroyed and there is
only a weak irregular excitation of spins between the skyrmions as
the system went off-resonance with microwaves. The overall view is
close to that of a thermally excited SkL shown in Fig. \ref{Fig_SkL_thermalized}.}

\label{Fig_LF_pumping_advanced}
\end{figure}
\begin{figure}
\centering{}\includegraphics[width=8cm]{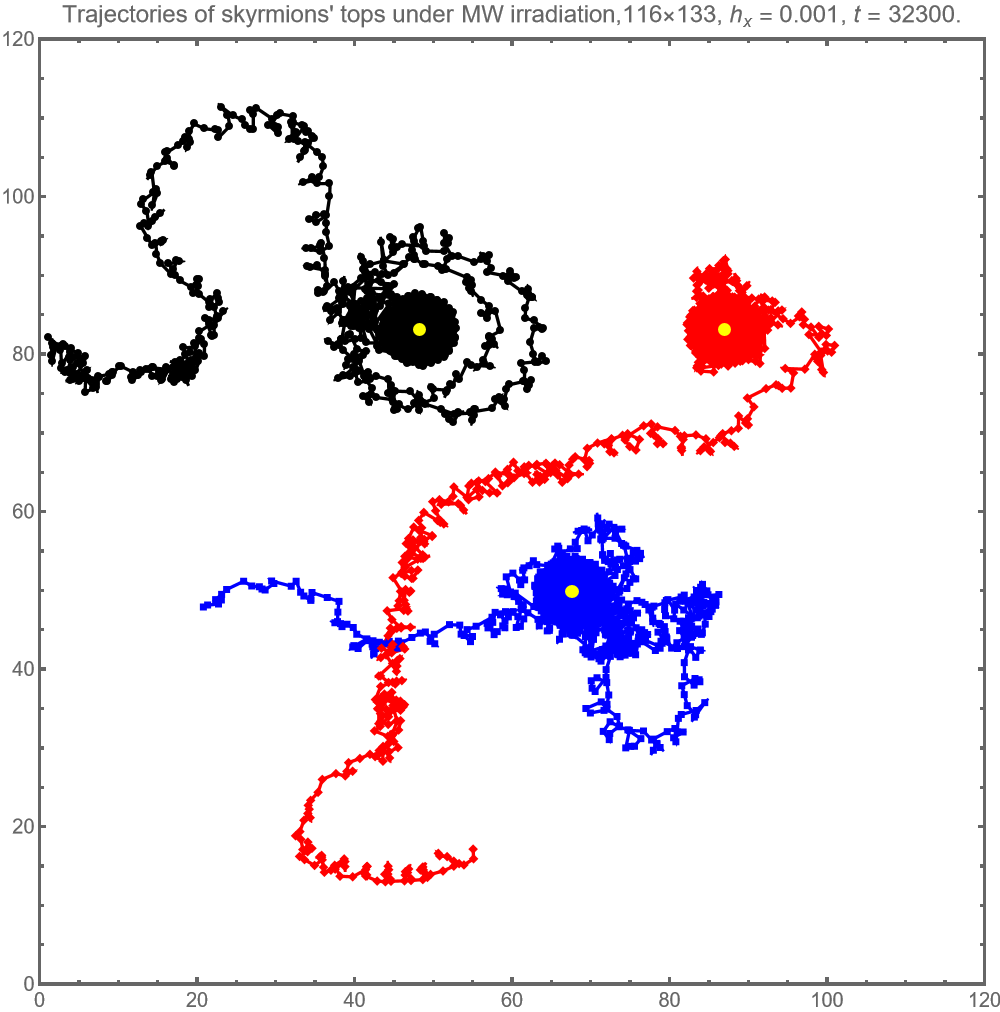}
\caption{Trajectories of the centers of three skyrmions shown in the pictures
above. For some time, the centers are rotating around their initial
positions, then they go astray and the SkL gets destroyed.}
\label{Fig_skyrmion_trajectories}
\end{figure}

Skyrmion lattice strongly interacts with microwaves at resonance with
one of its uniform modes. The most interesting is resonance pumping
of the LF mode by the linearly polarized microwaves with the magnetic-field
vector in the $xy$ plane which leads to the destruction of the SkL.
The same for the HF mode yields inconclusive results because in this
mode resonating spins occupy only a small part of the system and do
not affect the whole SkL much. On the other hand, resonance pumping
of the breathing mode by the microwaves with the magnetic field vector
along $z$-axis quickly leads to the collapse of skyrmions themselves
which is a single-skyrmion phenomenon and is thus less interesting.
So here we focus on the LF-mode pumping.

Different stages of irradiation of the SkL by MW of the amplitude
$h_{0x}/J=0.001$, resonating with the LF mode, are shown in Figs.
\ref{Fig_LF_pumping_early} and \ref{Fig_LF_pumping_advanced}, where
positions of three observed skyrmions are marked by black dots. At
the early stages shown in Fig. \ref{Fig_LF_pumping_early}, down-oriented
spins in the regions between skyrmions are precessing in-phase with
increasing and decreasing amplitude. As the precession amplitude becomes
large, skyrmions become noticeably deformed, losing their angular
symmetry. In this case, skyrmions centers found by the locator formula,
Eq. (\ref{Skyrmion_locator}), do not exactly coincide with their
tops. Skyrmions are rotating in-phase around their inital positions
in the SkL. At the advanced stages shown in Fig. \ref{Fig_LF_pumping_advanced},
precession of the down-oriented spins gradually loses its coherence
and skyrmions become deformed stronger. Then SkL gets destroyed and
apparently gets off-resonance with the MW as the precession of down-oriented
spins becomes less regular and weaker. This is off-resonance behavior
is natural as the frequencies of the SkL modes change with the thermal
excitation of the SkL, see Fig. \ref{Fig_Modes_frequencies_and_damping}.
SkL destroyed by resonant microwaves looks similar to the thermally-excited
SkL at $T/J=0.11$ in Fig. \ref{Fig_SkL_thermalized}. However, the
spin temperature at $t=17100$ in Fig. \ref{Fig_LF_pumping_advanced}
(lower) is only $T_{S}/J=0.00907$.

Fig. \ref{Fig_skyrmion_trajectories} shows trajectories of the centers
of the three observed skyrmions in the course of the resonant LF pumping.
At the early stages, the skyrmions are rotating around their initial
positions with the orbiting radius increasing and decreasing. Then
the skyrmions go astray, their motion becomes stochastic, and the
SkL gets destroyed.

\begin{figure}
\centering{}\includegraphics[width=8cm]{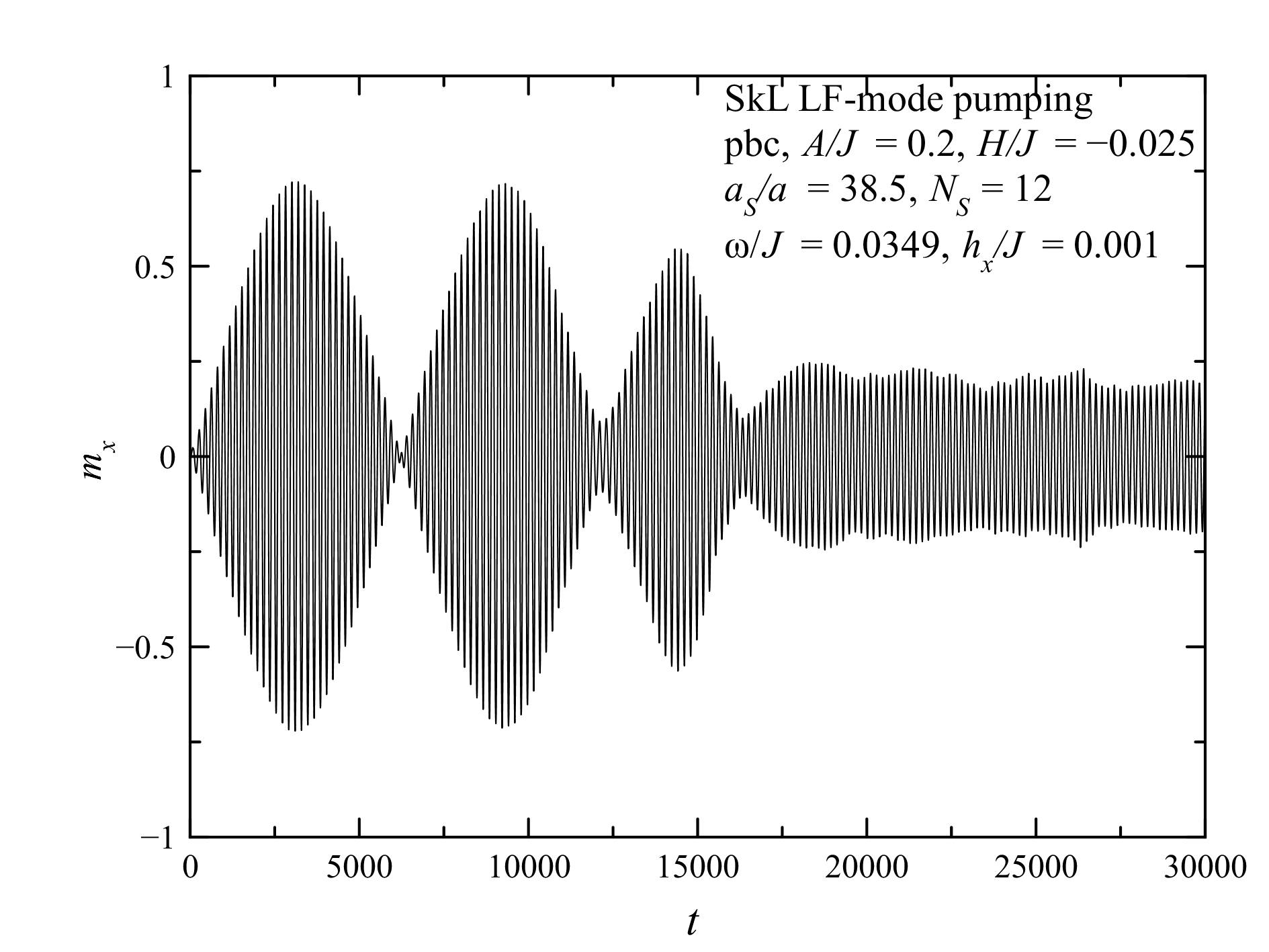} \caption{Time dependence of the transverse magnetization component in the course
of resonance pumping of the LF mode at $T=0$.}
\label{Fig_m_x_vs_t} 
\end{figure}

Time dependence of the transverse magnetization $m_{x}$ ($m_{y}$
behaves similarly) is shown in Fig. \ref{Fig_m_x_vs_t}. The main
frequency of this motion is the frequency of the LF mode at $T=0$,
i.e., $\omega_{xy}/J=0.0349$. The amplitude of this precession increases
and decreases a couple of times before it stabilizes at a low level
with small fluctuations. This phenomenon is a classical analog of
Rabi oscillations for a quantum two-level system which is typical
for non-linear classical systems, including spins and oscillators
(see, e.g., Ref. \cite{Claudon-PRB2008}). The system absorbs the
energy and then releases it, and the frequency of this process grows
with the amplitude of the pumping. This process is recurrent if there
is only one mode which cannot absorb the energy above a certain limit.
The single-mode excitation/deexcitation scenario hold here for some
time, and then processes of the energy transfer into other excitation
modes kick off. The spin precession stabilises at a lower level and
fluctuates because of the interaction with other modes. From these
other modes, the energy is migrating to further modes, including non-uniform
modes and a long cascade of transformations finally becomes thermal
energy. The presence of non-uniform modes manifests itself in the
breakdown of the coherent rotation of skyrmions and the destruction
of the SkL.

\begin{figure}
\begin{centering}
\includegraphics[width=8cm]{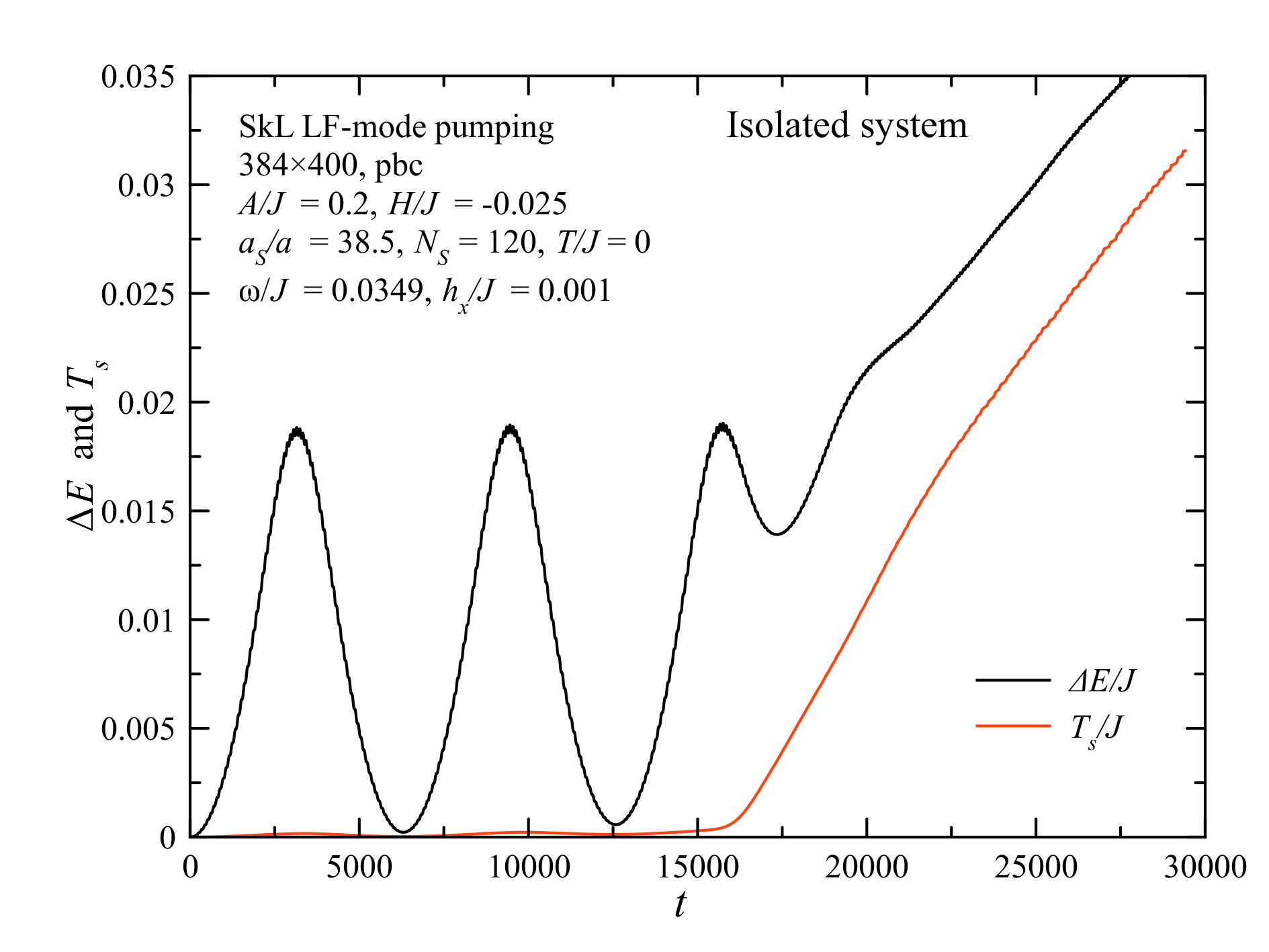} 
\par\end{centering}
\centering{}\includegraphics[width=8cm]{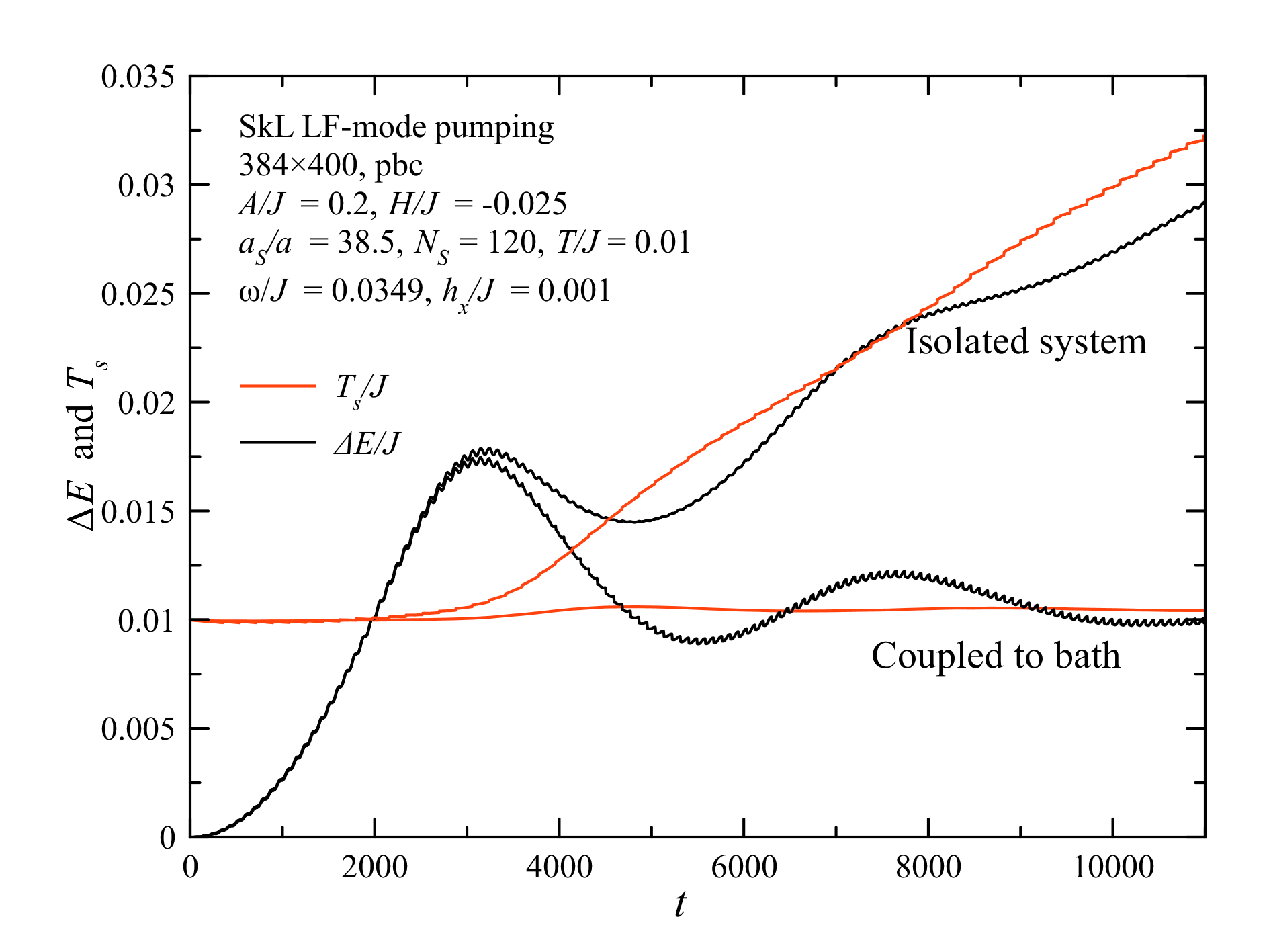}\caption{Time dependence of the system's energy and spin temperature in the
course of resonant pumping of the LF mode at $T=0$ and $T/J=0.01$,
with and without coupling to the bath.}
\label{Fig_E_Ts_vs_t} 
\end{figure}

It is enlightening to follow the time dependence of the energy of
the system and spin temperature in the cases of the isolated system
and the system coupled to the bath. The increase of the system's energy
$\Delta E$ per spin as the result of the MW irradiation and $T_{S}$for
an isolated system, initially at $T=0$, is shown in the upper panel
of Fig. \ref{Fig_E_Ts_vs_t}. In this case the energy was corrected
to the absorbed MW energy $E_{\mathrm{abs}}$, so that both energies
go together and coincide at the time points where the MW field $h(t)=0$.
As said above, the energy $\Delta E$ increases and decrease a couple
of times before the process breaks down because of the interaction
of the directly excited mode with secondary modes. Within this whole
time interval, the spin temperature $T_{S}$ remains very close to
zero in spite of the energy oscillations. This is because the dominant
contribution to $T_{S}$ in Eq. (\ref{TS}) comes from the short-wavelength
spin waves which play the role of the thermal reservoir of the system.
On the other hand, long-wavelength excitations directly excited in
this experiment, even at high amplitudes, do not significantly contribute
to $T_{S}$. When the regular excitation/deexcitation of the LF mode
breaks down at $t\simeq15000$, $T_{S}$ starts to grow steadily from
nearly zero, as the system continues to absorb energy and $\Delta E$
also steadily increases.

The lower panel of Fig. \ref{Fig_E_Ts_vs_t} shows the time dependence
of $\Delta E$ and $T_{S}$ for the isolated system prepared at $T/J=0.01$
and the system coupled to the bath with $T/J=0.01$. Here, there are
already thermal excitations in the initial state which interact with
the excited LF mode. This leads to a faster breakdown of the regular
excitation/deexcitation regime after which the energy and the temperature
of the isolated spin system steadily increase. The spin temperature
of the coupled system remains very close to the bath value, whereas
the energy does not increase but slowly oscillates with decreasing
amplitude. Small wiggles in the energy curve have the frequency of
the MW field, $\omega/J=0.0349$.

\begin{figure}
\begin{centering}
\includegraphics[width=8cm]{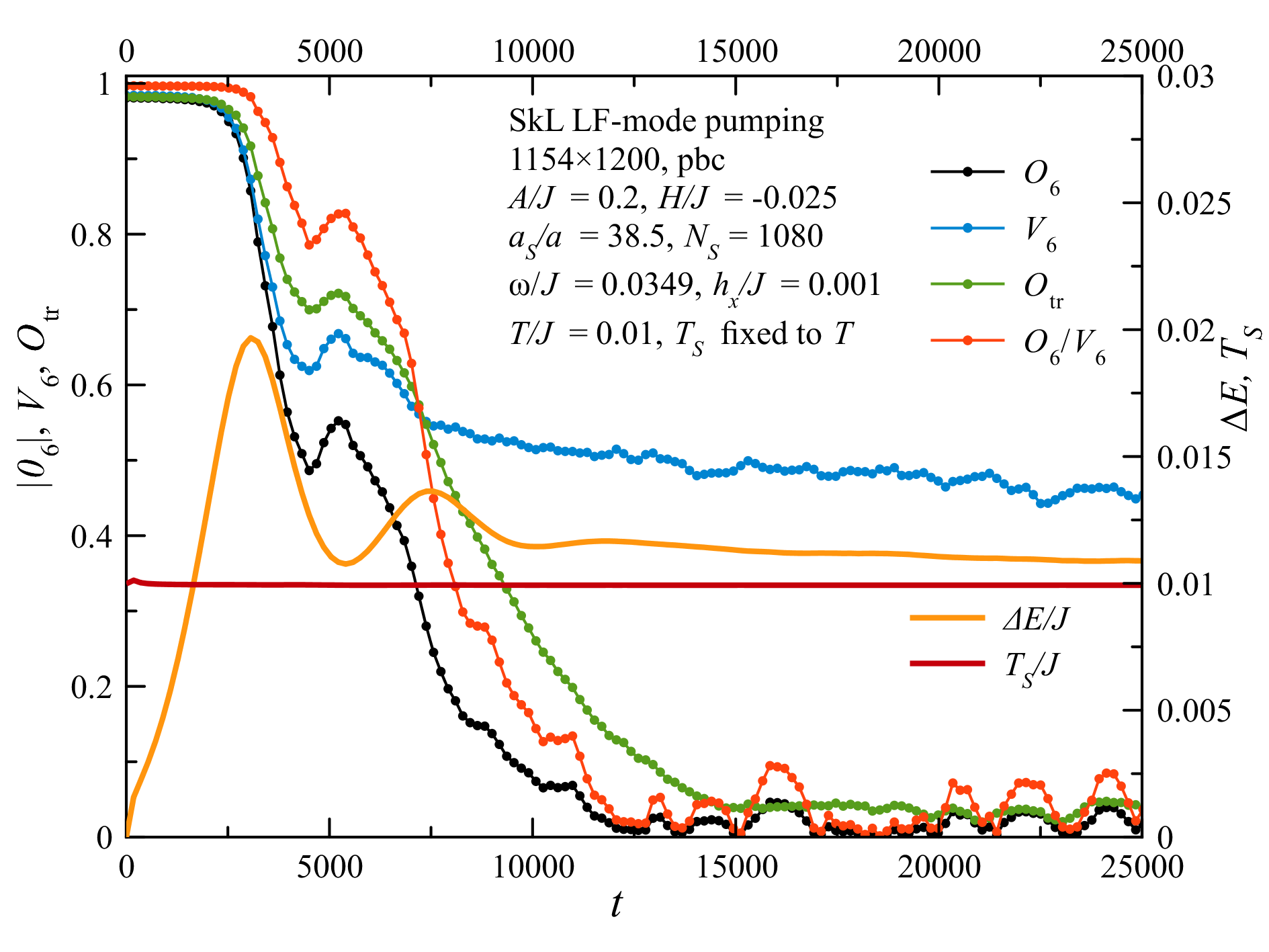} 
\par\end{centering}
\centering{}\includegraphics[width=8cm]{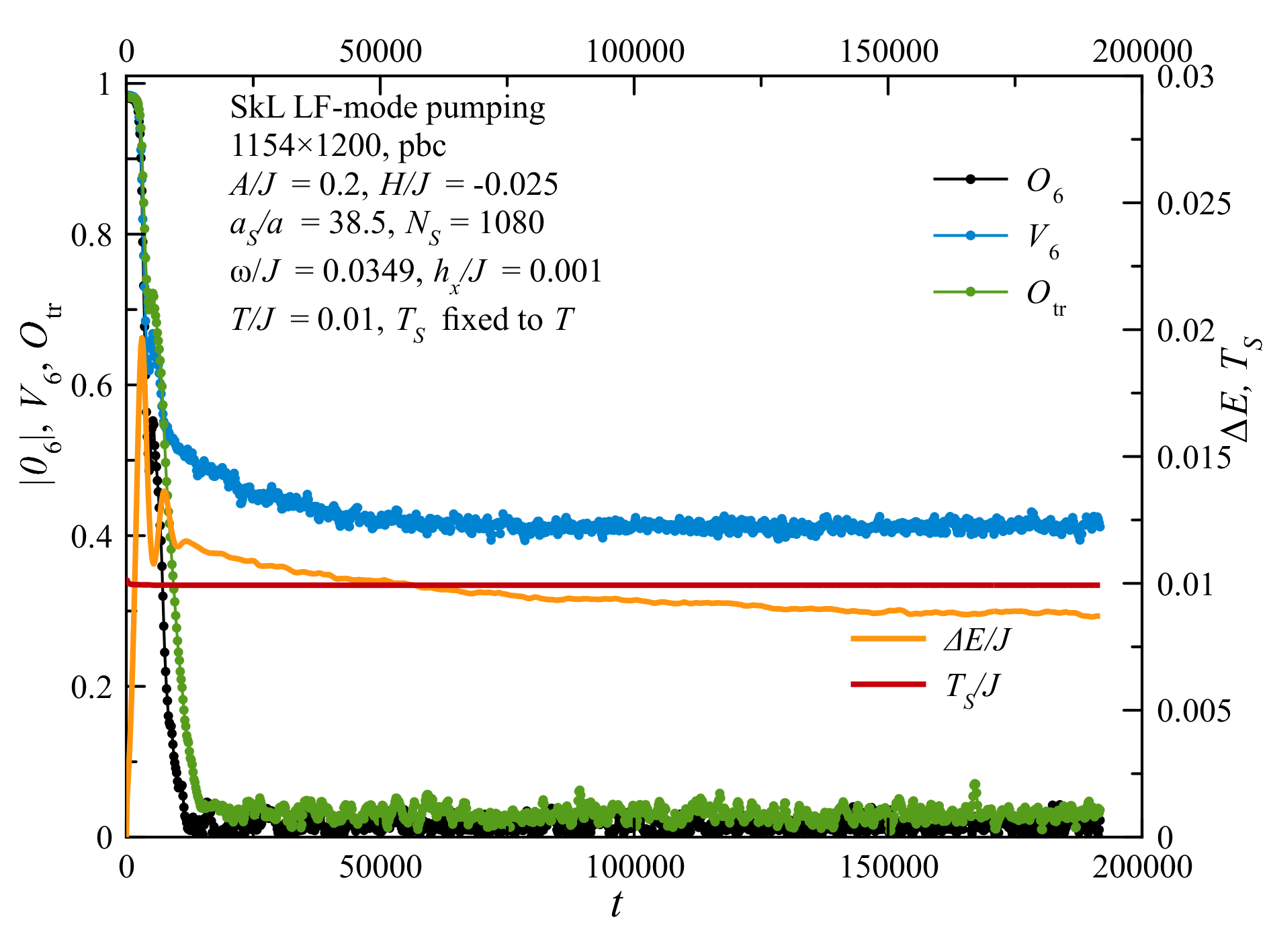}\caption{Time dependence of the parameters of the SkL, as well as of the energy
and spin temperature, in the course of resonance excitation of the
LF mode by microwaves with the amplitude $h_{x0}/J=0.001$. Upper
panel: short-time region. Lower panel: whole time region. One can
see that at long times the system reaches a stationary state in which
the SkL is destroyed, in spite of the temperature being kept low.}
\label{Fig_O6V6OtrTsDeltaE} 
\end{figure}
\begin{figure}
\begin{centering}
\includegraphics[width=8cm]{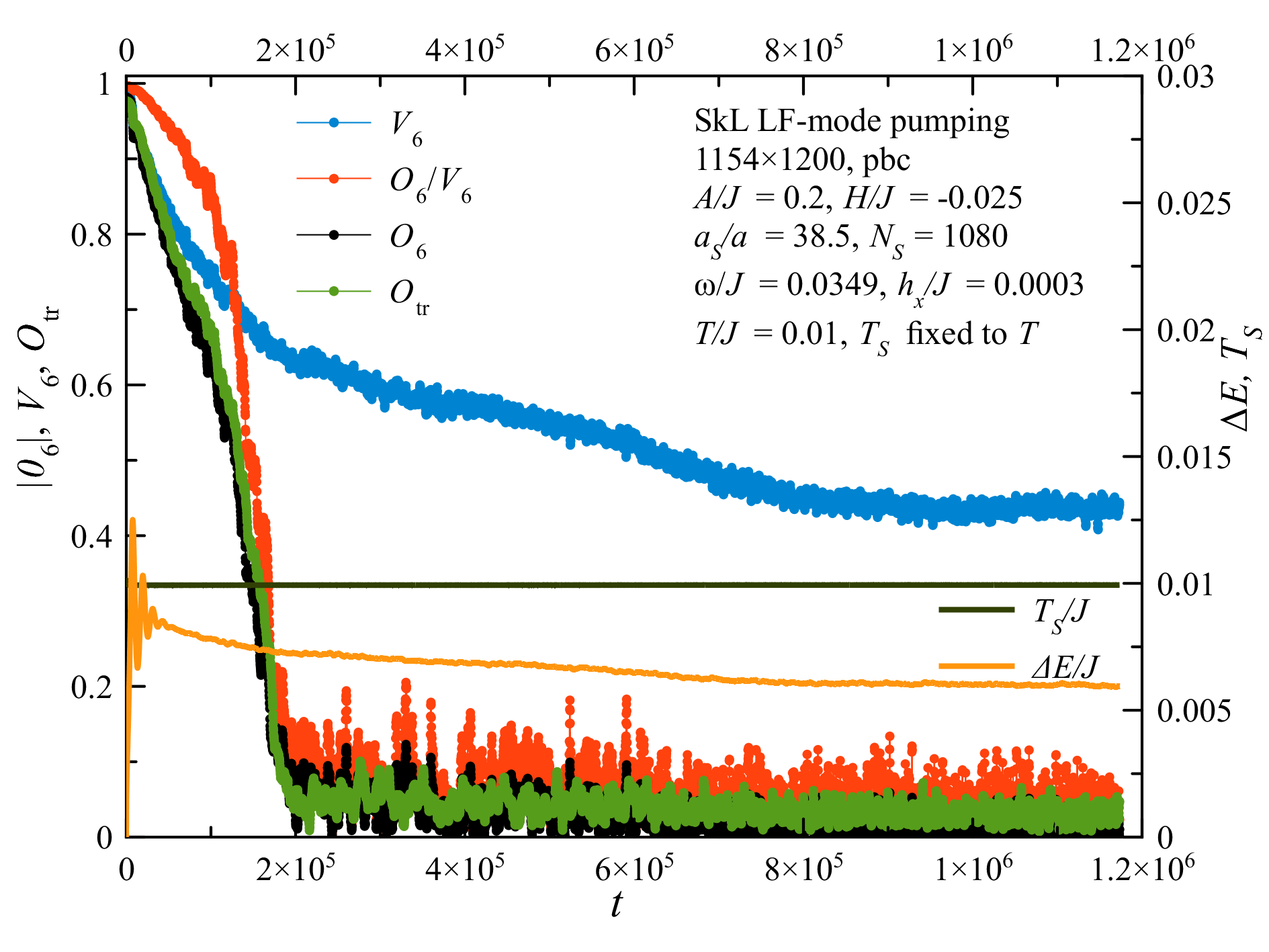}
\par\end{centering}
\caption{Time dependence of the parameters of the SkL, as well as of the energy
and spin temperature, in the course of resonance excitation of the
LF mode by microwaves with the smaller amplitude $h_{x0}/J=0.0003$.}

\label{Fig_O6V6OtrTsDeltaE_hx=00003D0.003}
\end{figure}
Let us now discuss the destruction of the skyrmion lattice by resonant
microwaves in terms of the SkL parameters introduced in Sec. \ref{Sec_SkL}.
Fig. \ref{Fig_O6V6OtrTsDeltaE} shows time dependences of $O_{6}$,
$V_{6}$, their ratio $O_{6}/V_{6}$, and $O_{tr}$, as well as the
system's excess energy $\Delta E$ and the spin temperature $T_{S}$
on the right $y$ axis, for the same MW amplitude $h_{x0}/J=0.001$
for the system coupled to the bath with the temperature $T/J=0.01$.
Since the results for small systems reported on above become here
rather noisy, we performed the computation on a larger system of $1154\times1200$
spins comprising $N_{S}=1080$ skyrmions, as in Ref. \cite{DG-EC-JMMM2024}.
The SkL parameters decrease with a faint recovery attempt at $t\approx5000$
(a reproducible feature) and the SkL gets finally destroyed at $t\approx10000$.
At the early stages of the MW irradiation, $t\lesssim2500$, skyrmions
rotate in-phase around their equilibrium positions, and SkL remains
intact. Comparing the time dependence of $O_{6}$ with that of the
ratio $O_{6}/V_{6}$ shows that the deterioration of the SkL is partially
due to that of the lattice hexagons (which reduces $V_{6}$) and that
of the long-distance correlation of the hexagons' orientations. At
long time the hexagon quality stabilizes at $V_{6}\simeq0.41$ which
corresponds to the complete hexagon disordering, $V_{6}=\sqrt{1/6}\simeq0.408$.

A similar experiment with a smaller MW amplitude $h_{x0}/J=0.0003$
shows in Fig. \ref{Fig_O6V6OtrTsDeltaE_hx=00003D0.003} a dramatic
slowing down of the process. Partially this can be explained by the
dependence of the microwave absorption $P\propto h_{0}^{2}$ which
accounts for a factor of about 10. However, here the SkL is destroyed
at $t\simeq2\times10^{5}$ which yields a slow-down by a factor of
20 in comparison to the results for $h_{x0}/J=0.001$. Thus, the experiment
took a very long computing time, and even at $t=10^{6}$ the hexagon
quality $V_{6}$ hasn't yet stabilized. The question arising here
is whether weaker microwaves would destroy the skyrmion lattice. For
an isolated system, sooner or later this will happen because of the
gradual warming of the system by the absorbed microwaves. However,
especially for weak microwaves, the coupling to the environment cannot
be neglected and it keeps the temperature practically constant. 

\section{Conclusions}

\label{Sec_conclusions}

We have studied excitation modes and the process of disordering of
skyrmion lattices subjected to the resonant microwave field. The three
modes we identified: the high- and low-frequency precessing modes
and the breathing mode are in agreement with the previously obtained
results \cite{Mochizuki-PRL2012,Onose-PRL2012,Aqeel-PRL2021,Lee-JPhys2022,Satywali-NatCom2021,Li-JPhys2023}.
The absorption of the microwave power by the high-frequency mode is
much weaker than by the low-frequency and breathing modes in the most
of the parameter range. We obtained the dependence of the frequencies
of these modes on temperature and magnetic field. The low-frequency
mode has an almost linear dependence on the magnetic field, while
the high-frequency mode and the breathing mode exhibit a non-monotonic
field dependence of the frequency, with a minimum in the intermediate
field range. The temperature dependence of the low-frequency mode
is expectedly decreasing, while that of the breathing mode is surprisingly
increasing. Both of these temperature dependences are close to linear
and do not show any features near the SkL melting temperature. 

When solving the dynamical equations for the spins numerically, we
did not use any phenomenological damping constants. The nonlinearity
of the system generates a substantial intrinsic damping that increases
on raising temperature. The damping of the low-frequency mode has
almost linear dependence on temperature, while that of the breathing
mode exhibits a stronger, nonlinear temperature dependence.

Pumping the microwave power at resonance with the LF mode into the
system begins with Rabi-like oscillations of the magnetization which
level off as the pumping continues. As in the case of quantum Rabi
oscillations, the frequency of these classical oscillations, previously
observed in a Josephson device modeled by an anharmonic oscillator
\cite{Claudon-PRB2008}, is determined by the amplitude of the ac
field. The absorption of microwave energy by spins increases the amplitude
of spin precession. As it reaches maximum the spins begin to emit
energy and decrease their precession amplitude. This process repeats
itself periodically, which explains the Rabi-like oscillations. As
the transfer of energy into other modes kicks in, these oscillations
become irregular and wash out.

We also performed detailed studies of the melting of the SkL by microwaves,
preceded by our studies of the melting of SkL on raising its temperature
\cite{GC-PRB2023,DG-EC-JMMM2024,DG-JS-EC-JPhys2024}. For the same
choice of parameters as here, the latter occurred at $T\approx0.12J$.
When pumping resonant microwaves, one can maintain the SkL at a constant
temperature by providing good thermal contact between the ferromagnetic
film and the substrate at a constant temperature. We have found that
the melting of the SkL by microwaves occurs regardless of temperature,
including temperatures as low as $T=0.01J$, indicating its non-thermal
character. Nevertheless, its general feature remains the same as the
melting on heating: a one-stage process characterized by the simultaneous
loss of the long-range translational and orientational order.

It would be interesting to study experimentally how fundamental this
property of the skyrmion crystal is and whether skyrmion lattices
in magnetic films with other kinds of interactions exhibit a different
behavior, such as, e.g., a two-stage melting predicted in \cite{HN-PRL1978,NH-PRB1979,Young-PRB1979}
and observed in colloidal crystals \cite{Gunberg-PRL2004,Zanghellini-2005}
where the orientational order exhibits more robustness than the translational
order. Melting of skyrmion lattices by microwaves may be a good avenue
for such studies.

\section{Acknowledgments}

This work has been supported by Grant No. FA9550-24-1-0090 funded
by the Air Force Office of Scientific Research.

\end{document}